\documentclass[12pt]{iopart}

\usepackage{iopams}
\usepackage{graphicx}
\usepackage{subfig}

\def\br{\begin{eqnarray}}
\def\er{\end{eqnarray}} 
\def\be{\begin{equation}}
\def\ee{\end{equation}}
\def\ba{\begin{array}}
\def\ea{\end{array}}
\def\bdm{\begin{displaymath}}
\def\edm{\end{displaymath}}
\def\nonu{\nonumber}
\def\lb{\left\lbrack}
\def\rb{\right\rbrack}
\def\({\left(}
\def\){\right)}

\def\a{\alpha}
\def\b{\beta}
\def\d{\delta}

\def\k{\kappa}
\def\l{\lambda}

\def\m{\mu}

\def\O{\Omega}
\def\p{\phi}

\def\t{\tau}

\def\Th{\Theta}

\def\cm{{\cal M}}
\def\ck{{\cal K}}

\def\hc{\hat{c}}
\def\hd{\hat{d}}

\def\pa{\partial}

\def\rlx{\relax\leavevmode}
\def\inbar{\vrule height1.5ex width.4pt depth0pt}
\def\IZ{\rlx\hbox{\sf Z\kern-.4em Z}}
\def\IR{\rlx\hbox{\rm I\kern-.18em R}}
\def\IC{\rlx\hbox{\,$\inbar\kern-.3em{\rm C}$}}
\def\one{\hbox{{1}\kern-.25em\hbox{l}}}
\def\soliton#1{$#1$--soliton}
\newcommand{\bra}[1]{\langle #1|}
\newcommand{\ket}[1]{|#1\rangle}

\newcommand{\expected}[1]{\langle #1 \rangle}

\def\NPB#1#2{{\it Nucl. Phys.} B {\bf #1} #2}

\def\PRE#1#2{{\it Phys. Rev.} E {\bf #1} #2}
\def\PLA#1#2{{\it Phys. Lett.} A {\bf #1} #2}

\def\JMP#1#2{{\it J. Math. Phys.} {\bf #1} #2}

\def\IJMPA#1#2{{\it Int. J. Mod. Phys.} A {\bf #1} #2}

\def\TMP#1#2{{\it Theor. Mat. Phys.} {\bf #1} #2}
\def\JPA#1#2{{\it J. Physics} A {\bf #1} #2}

\def\JPSJ#1#2{{\it J. of Phys. Soc. of Japan} {\bf #1} #2}

\def\CHAOS#1#2{{\it Chaos} {\bf #1} #2}

\begin{document}


\title[mKdV with nonvanishing boundary condition and the Gardner hierarchy]{
Nonvanishing boundary condition for the mKdV hierarchy and 
the Gardner equation}

\author{J F Gomes, Guilherme S Fran\c ca and A H Zimerman}

\address{
Instituto de F\'\i sica Te\' orica - IFT/UNESP\\
Rua Dr. Bento Teobaldo Ferraz, 271, Bloco II\\
01140-070, S\~ ao Paulo - SP, Brazil}

\ead{
\mailto{jfg@ift.unesp.br}, 
\mailto{guisf@ift.unesp.br}, 
\mailto{zimerman@ift.unesp.br}
}

\begin{abstract}
A Kac-Moody algebra construction for the integrable hierarchy 
containing the Gardner equation is proposed. 
Solutions  are systematically constructed  employing the dressing 
method and deformed vertex operators which takes into account the 
nonvanishing boundary value problem 
for the mKdV hierarchy. Explicit examples are given and besides
usual KdV like solitons, our solutions
contemplate the large amplitude table-top solitons, kinks,
dark solitons, breathers and wobbles.
\end{abstract}

\pacs{02.30.Ik, 02.30.Jr, 02.60.Lj, 05.45.Yv, 11.10.Lm, 47.35.Fg}

\submitto{J. Phys. A: Math. Gen.}

\maketitle

\section{Introduction}

The Gardner equation appeared a long time ago
when Miura \cite{miura1, miura2} introduced the remarkable 
transformation 
\be
u = v^2 + v_x ,
\label{miura}
\ee
connecting the Korteweg-de Vries (KdV) equation, 
$4 u_t = u_{3x} - 6 u u_x$, to the modified KdV (mKdV) equation,
$4 v_t = v_{3x} - 6 v^2 v_x$.
Both equations are ideal prototypes for integrable models.
The transformation \eref{miura} is highly nontrivial and relates 
solutions of the two nonlinear equations.

Miura \cite{miura1} pointed out that the KdV equation is invariant 
under Galilean transformation. In fact, if we change
the coordinates according to
\be
\label{galileu0}
x \to \xi + a\tau, \qquad t \to \tau, \qquad a=\mbox{const.},
\ee
then $\frac{\pa}{\pa x}=\frac{\pa}{\pa\xi}$,
$\frac{\pa}{\pa t}=\frac{\pa}{\pa\tau}-a\frac{\pa}{\pa\xi}$ and the KdV
equation becomes $4u_\tau-4au_\xi=u_{3\xi}-6uu_\xi$. The
undesirable $-4au_\xi$ term on the left hand side is canceled by the 
nonlinear term $-6uu_x$ on the right hand side, by a constant shift 
in the field variable. Therefore, the KdV equation is invariant under the 
transformation (\ref{galileu0}) with 
\br
u(x,t) \to u(\xi,\tau) + \case{2}{3}a.
\er
Following the same idea for the mKdV equation we get
$4v_\tau-4av_\xi=v_{3\xi}-6v^2v_\xi$. Because of the quadratic term in 
the nonlinearity, a constant shift in the field can still 
remove the undesirable $-4av_\xi$ term, but at the cost of an extra 
nonlinear one. Then, the mKdV equation is not invariant under the 
transformation (\ref{galileu0}) but it is transformed into the Gardner equation 
\be
4 v_\tau = v_{3\xi} - 6v^2v_\xi - 12\mu vv_\xi, \qquad 
\mu\equiv\(\case{2}{3}a\)^{1/2},
\label{gardner1}
\ee
if we shift the field 
\br
v(x,t) \to v(\xi,\tau) + \(\case{2}{3}a\)^{1/2}.
\label{galileu}
\er
\Eref{gardner1} is evidently interesting because it may be considered as 
combination of the KdV and mKdV equations or as an integrable deformation 
of the KdV or mKdV equations.
In a sequence of three papers, Wadati \cite{wadati1, wadati2, wadati3} 
obtained the \soliton{n} solution using the inverse scattering transform, 
B\" acklund transformation and showed that \eref{gardner1} is indeed a 
completely integrable Hamiltonian system.

It is known that integrable soliton equations appear as   
members of a more general structure called integrable hierarchies. 
In \cite{kuper} the results
previously obtained for the KdV equation individually
were generalized to the whole KdV hierarchy and its deformation
related to the Gardner transformation was discussed.

The properties of the Gardner equation is fairly well known and
can be obtained through the properties of the mKdV equation, and
is  subject of current research \cite{kiselev, munoz}. 
Besides the mathematical interest, equation \eref{gardner1} has a 
wide application 
in atmospheric and ocean waves and was extensively studied by Grimshaw
\etal \cite{grimshaw} where they obtained the large-amplitude 
table-top solitons, often observed in ocean coastal zones, and 
breathers \cite{grimshaw2}. 
Recently, solutions of the Gardner equation were used to construct 
static solutions of 
the Gross-Pitaevskii equation, known to describe dynamics of Bose-Einstein 
condensates \cite{malomed}.

A general algebraic approach for the construction of integrable 
hierarchies and its solitonic solutions can be formulated in terms
of representation of affine Lie algebras. The soliton  solutions  can be 
systematically constructed by the dressing method, which connects a 
trivial (vanishing) vacuum to a nontrivial soliton solution 
employing vertex operators. See for instance  
\cite{babelon, mira, jfg, nissimov, olive} and references therein.

The transformation \eref{galileu}, connecting solutions of the Gardner 
and mKdV equations, clearly shows that both solutions have different 
boundary values due to a constant shift in the field variable.  
This implies that  both equations  have different vacuum solutions.
In particular, those solutions of the Gardner hierarchy with trivial 
vacuum are related to nonvanishing vacuum solutions of  
mKdV equation and vice-versa.

In \cite{gui} we have extended the mKdV hierarchy for negative flows by 
introducing negative even graded Lax pairs. The nonlinear equations within 
that class do not present solutions with vanishing boundary condition
and, in order to circumvent this problem,  a deformed  
vertex operator was introduced. The present paper is a natural 
continuation of \cite{gui} in which 
the same algebraic formulation is employed to construct  
solutions of the Gardner equation. 
Interesting solutions arises
where a table-top soliton or a kink can 
coexist with normal solitons. We also obtain dark solitons, breathers and 
the wobble \cite{kalberman,luiz} solutions.

In sections 2 and 3  we briefly introduce the 
KdV and mKdV hierarchies, respectively, and show  
that Miura transformation represents a map between both hierarchies and not 
only between two single equations.
Our main results are contained in section 4 where we propose a deformed 
hierarchy grounded on $\hat{s\ell}_2$ Kac-Moody algebra, 
containing the Gardner equation as one of its members. 
In section 5 we introduce a new vertex operator and in 
section 6, construct explicit solutions for the whole hierarchy. 
Concluding remarks are in section 7.

\section{The KdV hierarchy}

Following \cite{miwa}, the mathematical construction of the KdV hierarchy 
can be introduced by the Lax equation in terms of pseudo-differential 
operators
\be
L_t = [L^{n/2}_{+}, L], \qquad n=1,3,5,\ldots
\label{kdv-hierarchy}
\ee
where $L\equiv\partial^2 - u(x,t)$ is the Schr\" odinger operator. 
The subscript $L_+$ denotes the differential part of the operator.
We then calculate
\br
L^{3/2}_{+} &= \partial^3 - \case{3}{2}u \partial - 
\case{3}{4}u_x, \label{pseudo3} \\
L^{5/2}_{+} &= \partial^5 - \case{5}{2}u \partial^3 - 
\case{15}{4}u_x \partial^2 + \(\case{15}{8}u^2-\case{25}{8}u_{2x}\)\partial -
\case{15}{16}u_{3x} + \case{15}{8}u u_x. \label{pseudo5}
\er
Substituting \eref{pseudo3} in \eref{kdv-hierarchy} we obtain 
the KdV equation
\be
4u_t = u_{3x} - 6uu_x.
\label{kdv}
\ee
Repeating the same calculation with \eref{pseudo5}, we have 
the Sawada-Kotera equation\footnote{The coefficients of this
5-th order KdV are different from the original Sawada-Kotera but the 
nonlinear terms are the same.}
\be
16u_t = u_{5x} - 10 u u_{3x} - 20 u_x u_{2x} + 30 u^2 u_x.
\label{sawada-kotera}
\ee

Higher order equations are obtained from \eref{kdv-hierarchy}
and its soliton solutions are well known \cite{miwa}.
It is important to note that while \eref{kdv} is invariant under
Galilean transformation, \eref{sawada-kotera} and higher
order KdV equations are not.

\section{The mKdV hierarchy}

The mKdV hierarchy can be constructed from a zero curvature condition
based on the affine Kac-Moody algebra $\hat{s\ell}_2$
\cite{ babelon, mira,  jfg, nissimov,  gui}, generated by
\be
\hat{s\ell}_2 = \{ E_{\a}^{(n)},\; E_{-\a}^{(n)},\; H^{(n)} \}
\ee
together with the spectral derivative operator $\hd$ and the central 
term $\hc$.
The $\hat{s\ell}_2$ commutation relations are
\br
\fl \lb H^{(n)}, H^{(m)} \rb = 2n\d_{n+m,0}\hc,\quad
\lb H^{(n)}, E_{\pm\a}^{(m)} \rb = \pm2 E^{(n+m)}_{\pm\a}, \quad
\lb E_{\pm\a}^{(n)}, E_{\pm\a}^{(m)}\rb = 0, \nonu \\
\fl \lb E_{\a}^{(n)}, E_{-\a}^{(m)} \rb = H^{(n+m)} + n\d_{n+m,0}\hc, \quad
\lb \hc, T^{(n)} \rb = 0, \quad 
\lb \hd, T^{(n)}\rb = nT^{(n)},
\label{commutation}
\er
where $T^{(n)}\in\hat{s\ell}_2$.
The operator $\hat{Q}=\case{1}{2}H^{(0)}+2\hd$ defines the grading operation
$[ \hat{Q}, T^{(n)} ]=n T^{(n)}$, that decomposes the algebra 
into even and odd graded subspaces 
\be
\hat{s\ell}^{(2n)}_2 = \{H^{(n)}\}, \quad
\hat{s\ell}^{(2n+1)}_2 = \{E_{\alpha}^{(n)},\; E_{-\alpha}^{(n+1)}\}, \quad
\hat{s\ell}^{(0)}_2 = \{H^{(0)}\}.
\label{grading}
\ee
Considering this grading structure, it follows from the Jacobi identity that
for $T^{(i)}\in \hat{s\ell}_2^{(i)}$ and
$T^{(j)}\in \hat{s\ell}_2^{(j)}$ we have 
$\lb T^{(i)}, T^{(j)}\rb \in \hat{s\ell}_2^{(i+j)}$.

The mKdV hierarchy is then defined by the zero curvature equation
\be
\lb \pa_x + E_\a^{(0)} + E_{-\a}^{(1)} + v H^{(0)},
    \pa_t + D^{(n)} + D^{(n-1)} + \cdots +D^{(0)} \rb = 0 
\label{mkdv-hierarchy}
\ee
where $n=1,3,5,\ldots$, $v=v(x,t)$ and 
$D^{(j)} \in \hat{s\ell}_2^{(j)}$. In the construction of integrable models
from the zero curvature equation, only the loop-algebra, 
that corresponds to set $\hc=0$ in the commutation relations 
\eref{commutation} is employed. \Eref{mkdv-hierarchy} is solved grade 
by grade, starting from the highest one, until the zero grade projection 
leads to the nonlinear time evolution equation. 
This procedure works in the following
way. Let $n=3$ for example, then from the grading structure \eref{grading} the 
operators involved in the construction \eref{mkdv-hierarchy} must be linear 
combination of the algebra generators in the form
\br
D^{(3)}&=a_3E_\alpha^{(1)} + b_3E_{\alpha}^{(2)}, \qquad
D^{(2)}=c_2H^{(1)}, \\
D^{(1)}&=a_1E_{\alpha}^{(0)}+b_1E_{\alpha}^{(1)},  \qquad
D^{(0)}=c_0H^{(0)}.
\er
The coefficients $(a_i,b_i,c_i)$ will be determined as functions of 
the field $v(x,t)$, by projecting the zero curvature equation 
\eref{mkdv-hierarchy} into each graded subspace. In this way, we obtain 
the nonlinear partial differential equation, from the zero grade projection, 
and also its Lax pair. Note that only the commutation relations 
\eref{commutation} (with $\hc=0$) are used in this calculation and no 
matrix representation is needed. So, for $n=3$ we get the well known
mKdV equation
\be
4v_t = v_{3x} - 6v^2 v_x
\label{mkdv}
\ee
while for $n=5$ we find
\be
16v_t = v_{5x} - 10v^2v_{3x} - 40vv_xv_{2x} - 10v_x^3 + 30v^4v_x.
\label{msawada-kotera}
\ee
Without loss of generality in the above calculations, all 
integration constants were chosen to vanish.

Using the Miura transformation \eref{miura} in \eref{kdv} 
and \eref{sawada-kotera}, we can relate them with the
corresponding equations in the mKdV hierarchy, \eref{mkdv} and
\eref{msawada-kotera}, respectively. This operation can be expressed
for both equations as
\be
\mbox{KdV}\( u\to v^2+v_x \) =  
\(2v + \pa_x\)\mbox{mKdV}\( v \).
\label{miura-maps}
\ee
After verifying this expression for this two particular cases, we now show 
that, in fact,  the Miura transformation holds for all higher orders,
i.e. the Miura transformation is a map between solutions of the 
entire mKdV hierarchy into solutions of the KdV hierarchy. 
Consecutive time evolutions of the KdV hierarchy are given by the recursion 
formula
\be
u_{t_{n+2}}=R u_{t_{n}}, \qquad u_{t_{1}} = u_x,
\label{kdv-recursion}
\ee
where $R \equiv \case{1}{4}\pa_x^2-u-\case{1}{2}u_x\pa_x^{-1}$. 
Substituting $u=v^2+v_x$ in both sides of \eref{kdv-recursion} we can write 
it in a factorized form as
\be
u_{t_{n+2}} = \(2v+\pa_x\)v_{t_{n+2}} = 
\(2v+\pa_x\)\(\case{1}{4}\pa_x^2-v^2-v_x\pa_x^{-1}v\)v_{t_{n}}.
\ee
Define now  $R' \equiv \case{1}{4}\pa_x^2-v^2-v_x\pa_x^{-1}v$, thus 
\be
v_{t_{n+2}} = R'v_{t_{n}}, \qquad v_{t_1} = v_x,  
\ee
which shows that entire  mKdV 
hierarchy is recursively generated by $R^{\prime}$  and 
\eref{miura-maps} holds for $n=1,3,5,\ldots$.

\section{A hierarchy containing the Gardner equation}

Motivated by the transformation \eref{galileu}, we now propose a deformation 
of the mKdV hierarchy \eref{mkdv-hierarchy},
\be\fl
\lb \pa_x + E_\a^{(0)} + E_{-\a}^{(1)} + 
(\mu+v)H^{(0)} ,
\pa_t + D^{(n)} + D^{(n-1)} + \cdots +D^{(0)} \rb = 0 
\label{gardner-hierarchy}
\ee
where $\mu=\mbox{const.}$, $v=v(x,t)$ and $n=1,3,5,\ldots$.  
In the usual algebraic construction \cite{jfg}, as used in 
\eref{mkdv-hierarchy}, 
the semi-simple element $E=E_\a^{(0)}+E_{-\a}^{(1)}$ is 
responsible for the algebra decomposition
$\hat{s\ell}_2 = \ck \oplus \cm$, its kernel $\ck$ and image $\cm$ subspaces.
In \eref{gardner-hierarchy} the deformed element 
$E_\a^{(0)}+E_{-\a}^{(1)}+\mu H^{(0)}$ contains both $\ck$ and $\cm$ 
components. Obviously, the  modification from \eref{mkdv-hierarchy} to 
\eref{gardner-hierarchy} corresponds to a simple translation in the 
field $v\to v+\mu$, however, it introduces important changes in 
constructing its solutions through the dressing method.
Reconsidering \eref{mkdv-hierarchy},  but now with a nonvanishing 
constant boundary condition, $v \to v_0$, the situation is 
quite the same as in \eref{gardner-hierarchy} with $v\to0$. Therefore, 
solutions with vanishing boundary condition of \eref{gardner-hierarchy} are  
related to solutions with nonvanishing boundary condition of 
\eref{mkdv-hierarchy}.

Solving \eref{gardner-hierarchy} for $n=3$, using the same procedure
described in the previous section for the mKdV hierarchy, we get
\be
v_t = \frac{1}{4}v_{3x} - \(\mu^2 - \frac{\alpha}{2}\)v_x - 3\mu v v_x - 
\frac{3}{2}v^2v_x
\ee
where $\alpha$ is an arbitrary integration constant.
Choosing $\alpha=2\mu^2$ we have the well known Gardner equation
\be
4v_t = v_{3x} - 12\mu v v_x - 6v^2 v_x
\label{gardner}
\ee
and its Lax pair\footnote{$L=\pa+A$, $L_t=\lb L,B \rb$.}
\numparts
\br
\fl A &= E_{\a}^{(0)} + E_{-\a}^{(1)} + \(\mu + v\)H^{(0)},
\label{potential1} \\
\fl B &= E_\a^{(1)} + E_{-\a}^{(2)} + \(\mu+v\)H^{(1)} + 
\case{1}{2}\(v_{x} - v^2 -2\mu v +2\mu^2\)E_{\a}^{(0)} - \nonu \\
\fl&-\case{1}{2}\(v_{x}+v^2 + 2\mu v - 2\mu^2\)E_{-\a}^{(1)} +
\case{1}{4}\(v_{2x}-2v^3 - 6\mu v^2 + 4\mu^3\)H^{(0)}.
\label{potential3}
\er
\endnumparts
Solving now \eref{gardner-hierarchy} for $n=5$ we obtain
\br
16v_t &=  
v_{5x} - 10v^2v_{3x} - 40vv_xv_{2x} - 10v_x^3 + 30v^4v_x -  \nonu \\
&- 20\mu vv_{3x} - 40\mu v_xv_{2x} + 120\mu^2 v^2 v_x + 120\mu v^3v_x.
\label{gardner5}
\er
The two arbitrary integration constants that show up are
conveniently chosen: $\alpha=4\mu^2$, $\beta=6\mu^4$. \Eref{gardner5} is
a combination of \eref{msawada-kotera} and \eref{sawada-kotera}.
Note that 
when $\mu\to 0$, \eref{gardner5} becomes \eref{msawada-kotera}
in the same way that \eref{gardner} becomes \eref{mkdv}. 
We point out that \eref{gardner5} is
not obtained from \eref{msawada-kotera} by the 
Galilean transformation \eref{galileu}. 
The Lax pair for \eref{gardner5} is
\numparts
\br
\fl A &= E_{\a}^{(0)} + E_{-\a}^{(1)} + \(\mu + v\)H^{(0)},
\label{potential5_0} \\
\fl B &= E_\a^{(2)} + E_{-\a}^{(3)} + \(\mu+v\)H^{(2)} +
\case{1}{2}\(v_{x} - v^2 -2\mu v +4\mu^2\)E_{\a}^{(1)} - \nonu \\
\fl&-\case{1}{2}\(v_{x}+v^2 + 2\mu v - 4\mu^2\)E_{-\a}^{(2)} + 
\case{1}{4}\(v_{2x}-2v^3 - 6\mu v^2 + 4\m^2 v + 8\mu^3\)H^{(1)} + \nonu \\
\fl&+\case{1}{8}\(v_{3x}-6v^2v_x-2vv_{2x}+v_x^2+3v^4-12\mu vv_x-2\mu v_{2x} +
\right. \nonu \\
\fl& \qquad\qquad\qquad \left. + 4\mu^2v_x + 12\mu v^3 + 8\mu^2v^2 - 
8\mu^3v+8\mu^4\)E_\a^{(0)} - \nonu \\
\fl&-\case{1}{8}\(v_{3x}-6v^2v_x+2vv_{2x}-v_x^2-3v^4-12\mu vv_x+2\mu v_{2x}+
\right. \nonu \\
\fl& \qquad\qquad\qquad \left. + 4\mu^2v_x-12\mu v^3-8\mu^2v^2+ 
8\mu^3v -8\mu^4\)E_{-\a}^{(1)} + \nonu \\
\fl&+\case{1}{16}\(v_{4x}-10v^2v_{2x}-10vv_x^2+6v^5-20\mu vv_{2x}-10\mu v_x^2+
\right. \nonu \\
\fl& \qquad\qquad\qquad \left. +30\mu v^4 + 40\mu^2 v^3 +16 \mu^5 \)H^{(0)}.
\label{potential5}
\er
\endnumparts
The hierarchy defined in \eref{gardner-hierarchy} can also be recursively 
generated through the following pseudo-differential operator
\br
v_{t_{n+2}} = Rv_{t_n}, \qquad v_{t_1}=v_x, \\
R \equiv \case{1}{4}\pa_x^2-\(v+2\mu\)v - v_x\(\mu\pa_x^{-1}+\pa_x^{-1}v\).
\er

\section{Vertex operator}

The usual $\hat{s\ell}_2$ vertex operator with principal gradation 
\be
V\(\k\) = \sum_{n=-\infty}^{\infty}\k^{-2n}
\lb H^{(n)}-\case{1}{2}\delta_{n0}\hc + 
\k^{-1}E_\a^{(n)}-\k^{-1}E_{-\a}^{(n+1)} \rb,
\label{usual-vertex}
\ee
solves the mKdV and sinh-Gordon equations with vanishing boundary 
condition. Here we introduce a modified  vertex operator that 
generalizes and takes into account  the nonvanishing boundary value 
problem for the mKdV hierarchy and henceforth, the Gardner hierarchy 
with vanishing boundary condition.

Define the following vertex operator depending on the 
parameters $(\k_i,\mu)$
\be
\fl V_i \equiv \sum_{n=-\infty}^{\infty}
\(\k_i^2-\mu^2\)^{-n}
\lb H^{(n)}+ \frac{\mu-\k_i}{2\k_i}\d_{n0}\hc +
\frac{1}{\k_i+\mu}E_{\a}^{(n)}-
\frac{1}{\k_i-\mu}E_{-\a}^{(n+1)}\rb.
\label{vertex} 
\ee
This vertex was proposed in \cite{gui} to solve the
negative even grade part of the mKdV hierarchy.
Note that when $\mu\to0$  the 
vertex \eref{usual-vertex} is recovered. 
The parameter $\mu$, also present in equation \eref{gardner}, is related to 
the nonvanishing boundary condition of
the mKdV hierarchy through the identification 
$\mu \leftrightarrow v_0$, where $v \to v_0$ is the constant value of the
field in $|x|\to\infty$.
Consider the operator
\be
\O_m \equiv E_\a^{(m)} + E_{-\a}^{(m+1)} + \mu H^{(m)}
\label{op_m}
\ee
which corresponds to the vacuum $v=0$ configuration of the 
Lax component $A_{vac} = A_0$ in \eref{potential1}, for $m=0$.
Taking the commutator of \eref{op_m} with the vertex \eref{vertex} we 
verify that
\be
\lb \O_m, V_i \rb = -2\k_i\(\k_i^2-\mu^2\)^m V_i \, .
\label{eigenvalue}
\ee
As will be clear in the next section, \eref{eigenvalue}
determines the dispersion relation for all nonlinear evolution equations in 
the hierarchy.
Consider the highest weight states for $\hat{s\ell}_2$ ---
$\{\ket{\l_0},\;\ket{\l_1}\}$ --- which obey the following actions:
$E^{(0)}_{\a}\ket{\l_j} = 0$, $E^{(n)}_{\pm\a}\ket{\l_j} = 0$ and
$H^{(n)}\ket{\l_j} = 0$ for $n > 0$, $H^{(0)}\ket{\l_j} = \d_{j1}\ket{\l_j}$
and $\hc\ket{\l_j} = \ket{\l_j}$, for $j=0,1$. The adjoint relations are: 
${H^{(n)}}^{\dagger}=H^{(-n)}$, 
${E_{\alpha}^{(n)}}^{\dagger} = E_{-\alpha}^{(-n)}$ and 
$\hc^{\dagger}=\hc$. Taking \eref{vertex} between the highest weight states,
\be
\bra{\l_j}V_i\ket{\l_j} = \frac{\mu+\sigma_j\k_i}{2\k_i} \quad
\mbox{where} \quad
\sigma_0 = -1, \; \sigma_1 = 1 .
\label{coeff1}
\ee
A more involved calculation of two vertices yields
\be
\fl
\bra{\l_j}V_i V_l\ket{\l_j} = 
\frac{\(\mu+\sigma_j\k_i\)\(\mu+\sigma_j\k_l\)}{4\k_i\k_l}a_{ij}
\quad \mbox{where} \quad
a_{ij}\equiv\(\frac{\k_i-\k_l}{\k_i+\k_l}\)^2 .
\label{coeff2}
\ee
It is possible to prove that the expectation value of
a product of $n$ vertices decomposes into
\be
\bra{\l_j} \prod_{i=1}^{n}V_i \ket{\l_j} =
\prod_{i=1}^{n}\bra{\l_j}V_i\ket{\l_j}
\prod_{i,k=1,\,i<k}^{n}a_{ik}.
\label{coeffn}
\ee
The matrix elements \eref{coeff2} and \eref{coeffn} 
determine the nonlinear interaction between solitons.
Notice that the nilpotency property of the vertex representation 
is a direct consequence of \eref{coeff2} when $\k_i = \k_l$.
This corresponds to the physical interpretation that 
when $\k_i \to \k_l$ two solitons degenerate into a single soliton 
(no interaction). \Eref{coeffn} also corresponds the physical 
property that $n$ solitons interact in pairs.

\section{Dressing the vacuum}

The two dressing group elements $\Th_{\pm}$ correspond to gauge 
transformations mapping trivial vacuum potentials, $A_0$ and $B_0$, 
into nontrivial ones 
involving field dependent potentials, $A$ and $B$ \cite{nissimov, mira}
\numparts
\br
A &= \Th_\pm A_0 \Th_\pm^{-1}-\pa_x\Th_\pm\Th_\pm^{-1}, \label{gauge1} \\
B &= \Th_\pm B_0 \Th_\pm^{-1}-\pa_t\Th_\pm\Th_\pm^{-1}. \label{gauge2}
\er
\endnumparts
They are factorized into positive and negative grade generators
\be
\Th_+ = e^{X^{(0)}}e^{{X^{(1)}}}\ldots \qquad
\Th_- = e^{X^{(-1)}}e^{{X^{(-2)}}}\ldots \label{thetas}
\ee
which, together with the zero curvature condition yields,
\be
\Th_-^{-1}\Th_+ = \Psi_0 G \Psi_0^{-1}
\label{dressing}
\ee
where $G$ is an arbitrary constant group element and $\Psi_0$ is such that
$A_0=-\pa_x\Psi_0\Psi_0^{-1}$ and
$B_0=-\pa_t\Psi_0\Psi_0^{-1}$.
For the Gardner equation \eref{gardner},  
taking in \eref{potential1} and \eref{potential3}
we end up with  the vacuum potentials
$A_0 = \O_0$ and $B_0 = \O_1 + \mu^2 \O_0$, therefore
\be
\Psi_0 = \exp\lb -\O_0 x - \(\O_1 + \mu^2 \O_0\) t\rb
\label{psi_0_gardner}
\ee
where $\Omega_m$ is defined in \eref{op_m}.
The vacuum takes into account information about the boundary condition 
at $|x| \rightarrow \infty$ ($v \to 0$).

The dressing method requires the use of highest weight states representation
so it is necessary to include the central term $\hc$ in the 
calculations. In the construction of the models, because $\hc$ commutes
with every other operator, the zero curvature equation is invariant under
addition of a central term. Therefore, we change $A \to A -  \nu_x\hc$, where
$A=E^{(0)}_\alpha+E^{(1)}_{-\alpha}+vH^{(0)}$ is one of the Lax operators
and $\nu$ is a function that will be determined.
Solving for the zero grade projection of \eref{gauge1} with $\Th_+$ we find
\be
\exp\lb X^{(0)}\rb = \exp\lb \phi H^{(0)}+\nu\hc\rb, \qquad v=-\phi_x.
\label{x_0}
\ee
Using this result when taking 
the left hand side of \eref{dressing} between highest weight states,
\be
\fl
e^{\nu}=\bra{\l_0} \Th_-^{-1}\Th_+\ket{\l_0}\equiv\tau_0,\quad
e^{\p+\nu} = \bra{\l_1} \Th_-^{-1}\Th_+\ket{\l_1}\equiv\tau_1,\quad
v = \pa_x\ln\frac{\tau_0}{\tau_1}.
\label{solution}
\ee
Choosing now $G=\prod_{i=1}^{n}e^{V_i}$ and projecting the 
right hand side of \eref{dressing} between these states
\numparts
\br
\tau_j &=  
\bra{\l_j} \Psi_0\lb\prod_{i=1}^{n}\exp\{V_i\}\rb\Psi_0^{-1}\ket{\l_j} \nonu \\
&= \bra{\l_j}\prod_{i=1}^{n}\exp\{\Psi_0V_i\Psi_0^{-1}\}\ket{\l_j} \nonu \\
&= \bra{\l_j}\prod_{i=1}^{n}\exp\{e^{\xi_i(x,t)} V_i\}\ket{\l_j}
\label{lastone} \\
&= \bra{\l_j}\prod_{i=1}^{n}\(1+ e^{\xi_i(x,t)} V_i\)\ket{\l_j}
\label{tauj}
\er
\endnumparts
where in \eref{lastone} we have used the fact that $V_i = V(\k_i)$ 
are eigenstates of $\O_m$ with eigenvalues given by 
\eref{eigenvalue}.  It therefore follows 
the dispersion relation,
\be
\xi_i = 2\k_ix+\left\{2\k_i\(\k_i^2-\mu^2\)+2\k_i\mu^2\right\} t 
 = 2\k_ix+ 2\k_i^3 t
\label{spacetime}
\ee
where in \eref{tauj} the nilpotency property of the vertex $V_i$ was enforced.
The general formula \eref{tauj} is valid for all equations within the 
hierarchy and can be solved explicitly
by using the factorization property of the vertices \eref{coeffn} 
and \eref{coeff2}.
The explicit space-time dependence, however, is specified according to the 
choice of vacuum potentials $A_0 $ and $B_0$ for each individual model. 
Let us consider for instance the vacuum for potentials 
\eref{potential5_0} and \eref{potential5},
$A_0=\O_0$ and $B_0=\O_2+2\mu^2\O_1+\mu^4\O_0$. From \eref{eigenvalue} 
we therefore have
\br
\xi_i & = 2\k_ix+\left\{ 2\k_i\(\k_i^2-\mu^2\)^2+4\mu^2\k_i\(\k_i^2-\mu^2\)
+2\mu^4\k_i\right\} t \nonu \\
& = 2\k_i x + 2\k_i^5 t.
\label{spacetime2}
\er

\subsection{Nonvanishing boundary condition for the mKdV hierarchy}

We have constructed solutions to the hierarchy associated 
with the Gardner equation where  $v \to 0$ when $|x|\to\infty$. 
We now address the nonvaninshing boundary value
problem for the mKdV hierarchy.

Recall that in \eref{gardner-hierarchy} if we
take $\mu=0$ we recover the mKdV hierarchy \eref{mkdv-hierarchy}. 
When taking a constant
vacuum $v\to v_0$ in \eref{mkdv-hierarchy}, the term $v_0$ 
plays the role of $\mu$, and hence throughout this subsection
you shall consider all expressions after \eref{vertex} with $\mu$ replaced
by $v_0$.

Reconsider the zero grade projection of \eref{gauge1} 
for $\Theta_+$, but now with $A_0=E_\a^{(0)}+E_{-\a}^{(1)}+v_0H^{(0)}$ and 
$A=E_\a^{(0)}+E_{-\a}^{(1)}+v(x,t)H^{(0)}-\nu_x\hc$. 
The analogue of \eref{x_0} becomes
\be
\exp\lb X^{(0)} \rb = \exp \lb \(v_0 x+\p\)H^{(0)} + \nu\hc \rb, \qquad
v=-\phi_x,
\ee
and \eref{solution} implies the following nonvanishing boundary solution
\be
v = v_0 + \pa_x\ln\frac{\tau_0}{\tau_1}.
\label{mkdv-solution}
\ee
\Eref{tauj} does not change except for the dispersion relation. The vacuum
potentials are obtained by setting $v\to v_0$ in \eref{mkdv-hierarchy}. 
For the mKdV equation \eref{mkdv}
we have $A_0=\O_0$, $B_0=\O_1-\case{1}{2}v_0\O_0$ and for 
\eref{msawada-kotera} $A_0=\O_0$,
$B_0=\O_2-\case{1}{2}v_0^2\O_1+\case{3}{8}v_0^4\O_0$. Then, using 
\eref{eigenvalue}, we have the respective dispersion relations
\br
\xi_i &= 2\k_ix + \(2\k_i^3-3v_0^2\k_i\)t,
\label{spacetime-mkdv} \\
\xi_i &= 2\k_ix + \(2\k_i^5-5v_0^2\k_i^3+\case{15}{4}v_0^4\k_i\)t.
\label{spacetime-mkdv5}
\er
Note that the dispersion relations depend on the boundary condition, so
the velocity of this solitons depend on the vacuum field $v_0$.

\subsection{Examples}

We now present some explicit examples of the exact solutions we have obtained.
The tau functions for a general \soliton{n} solution  is given by \eref{tauj}.
The solution  is then obtained from 
 \eref{solution} or \eref{mkdv-solution}, with  space-time 
dependence specified  by the dispersion relations 
\eref{spacetime}--\eref{spacetime2} or 
\eref{spacetime-mkdv}--\eref{spacetime-mkdv5}, according to the Gardner or 
mKdV hierarchies, respectively.
 
The general expression for \soliton{1} solution is
\be
\tau_j = 1+\expected{V_1}_j e^{\xi_1}
\label{1soliton}
\ee
where $j=0,1$ and $\expected{\bullet}_j\equiv\bra{\l_j}\bullet\ket{\l_j}$.
The vertex element is given by \eref{coeff1} with $\mu \to v_0$ for the 
mKdV hierarchy.
The $2$-- and \soliton{3} solutions are respectively given by 
\be
\tau_j = 1 + \expected{V_1}_j e^{\xi_1} +
\expected{V_2}_j e^{\xi_2} +
\expected{V_1V_2}_j e^{\xi_1+\xi_2}
\label{2soliton}
\ee
and 
\br
\tau_j &= 1 +
\expected{V_1}_j e^{\xi_1} +
\expected{V_2}_j e^{\xi_2} +
\expected{V_3}_j e^{\xi_3} +
\expected{V_1V_2}_j e^{\xi_1+\xi_2} + \nonu \\
&+\expected{V_1V_3}_j e^{\xi_1+\xi_3} +
\expected{V_2V_3}_j e^{\xi_2+\xi_3} + 
\expected{V_1V_2}_j\expected{V_1V_3}_j\expected{V_2V_3}_j 
e^{\xi_1+\xi_2+\xi_3}.
\label{3soliton}
\er

\begin{figure}
\centering
\subfloat[]{\label{fig_1a}
\includegraphics[width=0.3\textwidth]{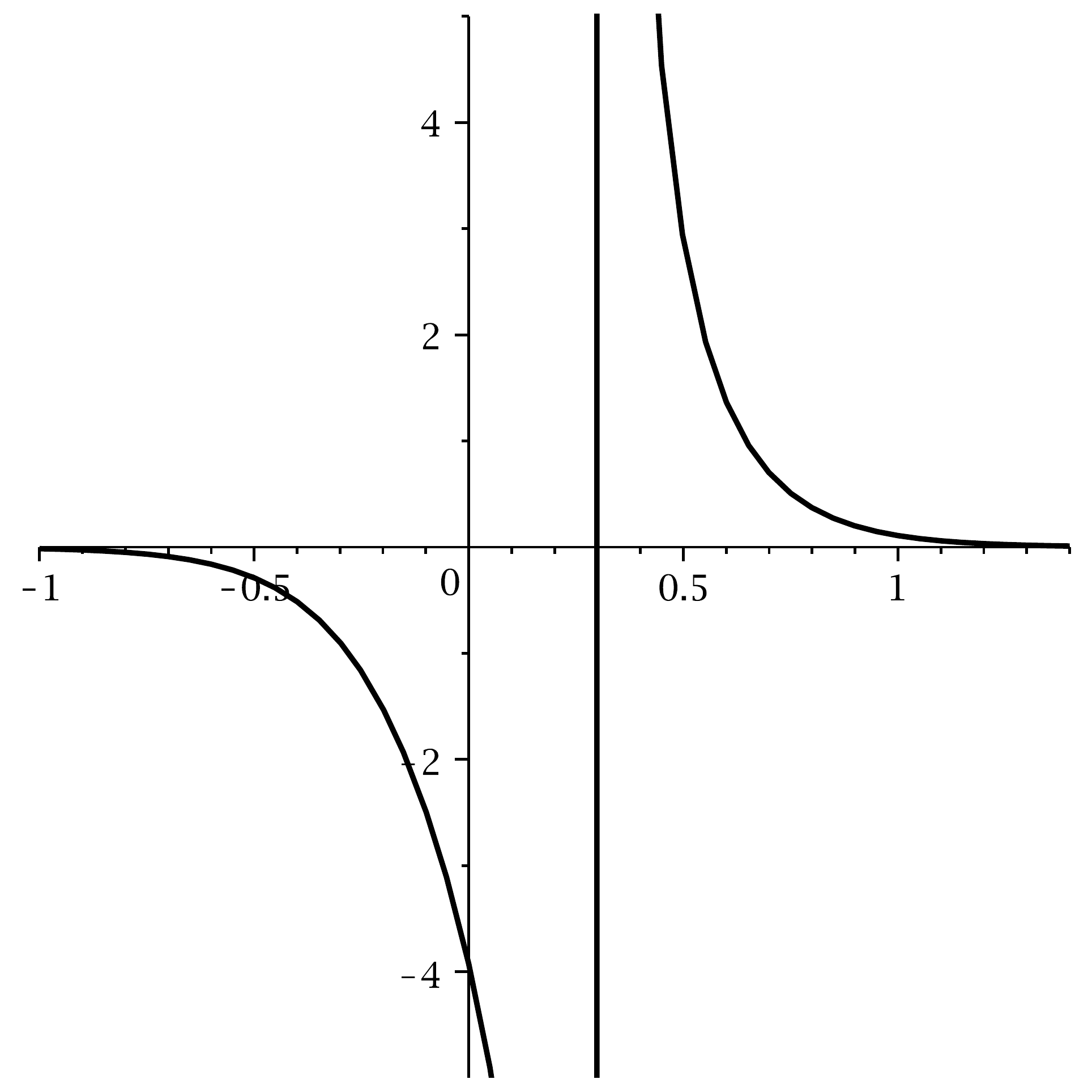}}
\subfloat[]{\label{fig_1b}
\includegraphics[width=0.3\textwidth]{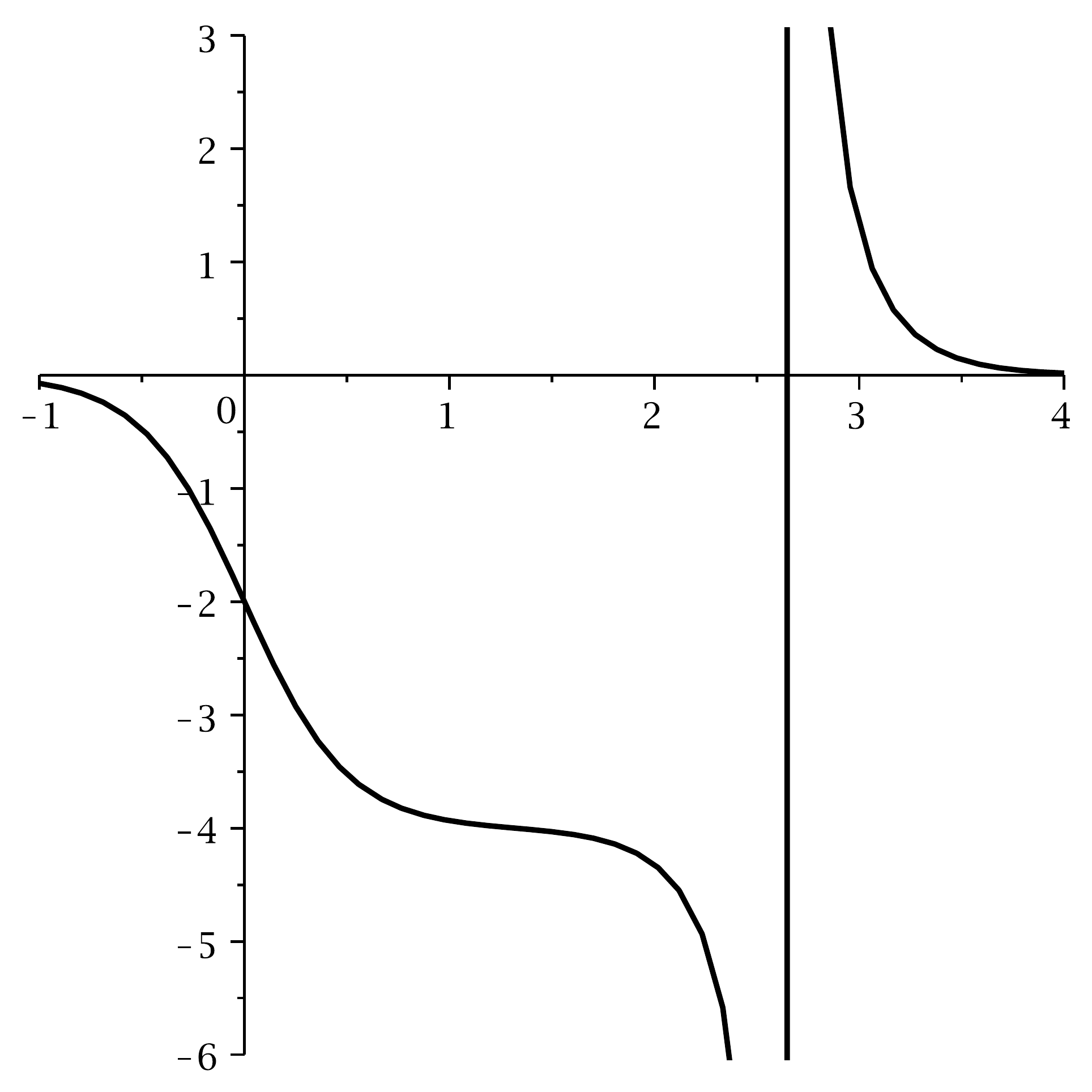}}
\subfloat[]{\label{fig_1c}
\includegraphics[width=0.3\textwidth]{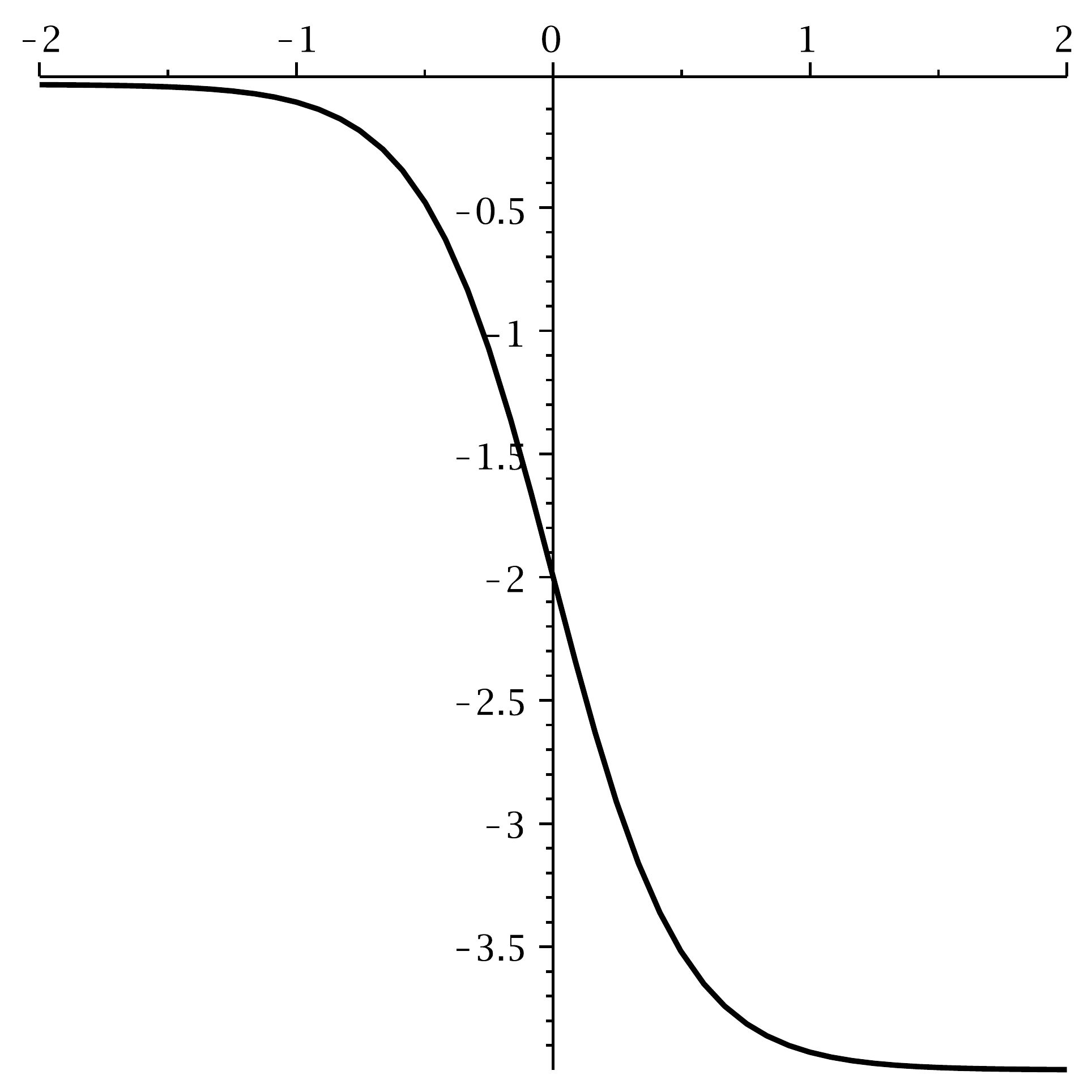}}
\caption{Soliton for the Gardner equation when $\k_1 \ge \mu $ and
$\mu > 0$. We choose $\mu=2$ and $\k_1 = (3,\ 2.0001,\ 2)$ in 
figures (a, b, c), respectively.
When $\k_1 \to \mu^{+}$ the solution shown in (a) 
deforms, as shown in (b), until it becomes a kink when $\k_1=\mu$ as
shown in (c).}
\label{fig_1}
\end{figure}

\begin{figure}
\centering
\subfloat[]{\label{fig_2a}
\includegraphics[width=0.3\textwidth]{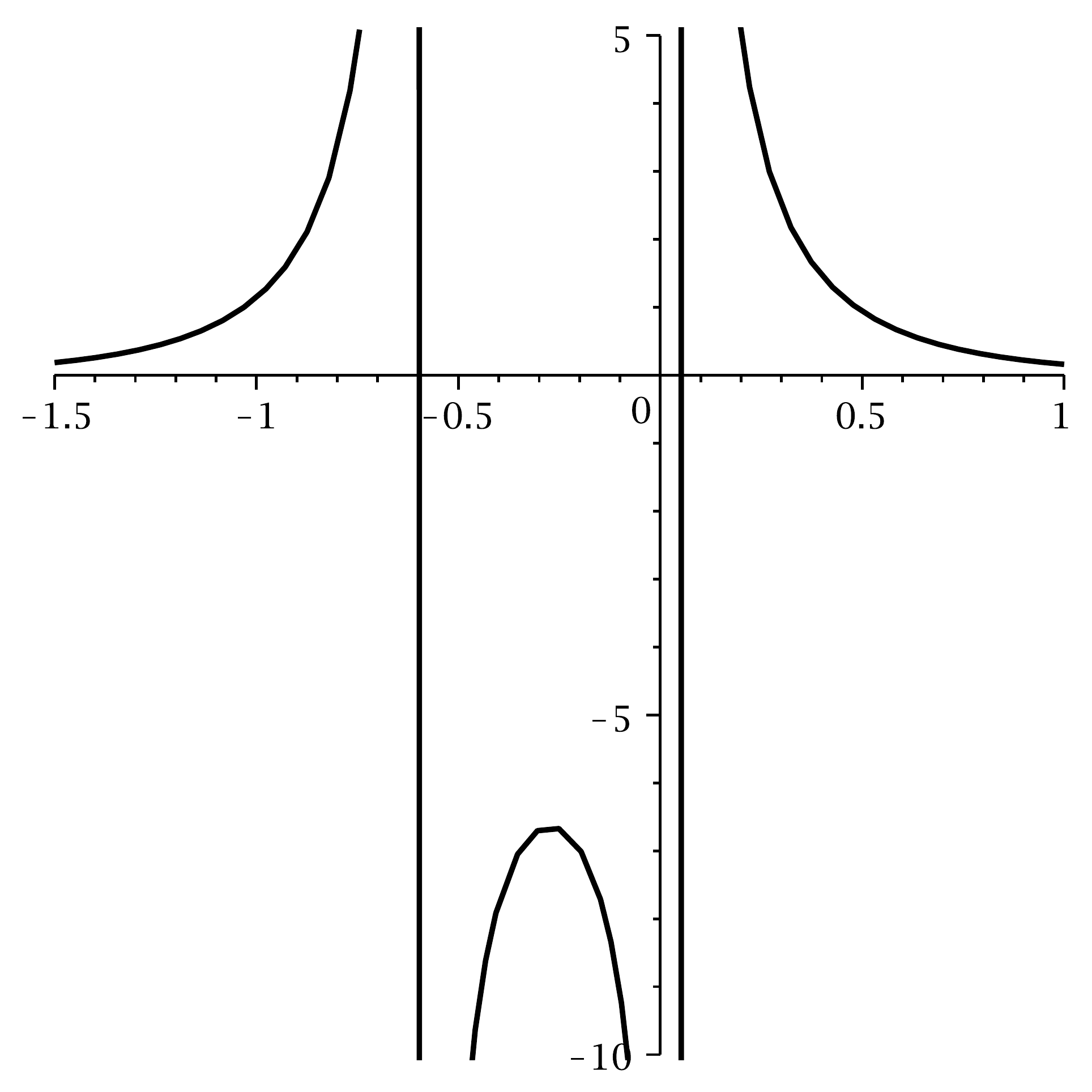}}
\subfloat[]{\label{fig:2b}
\includegraphics[width=0.3\textwidth]{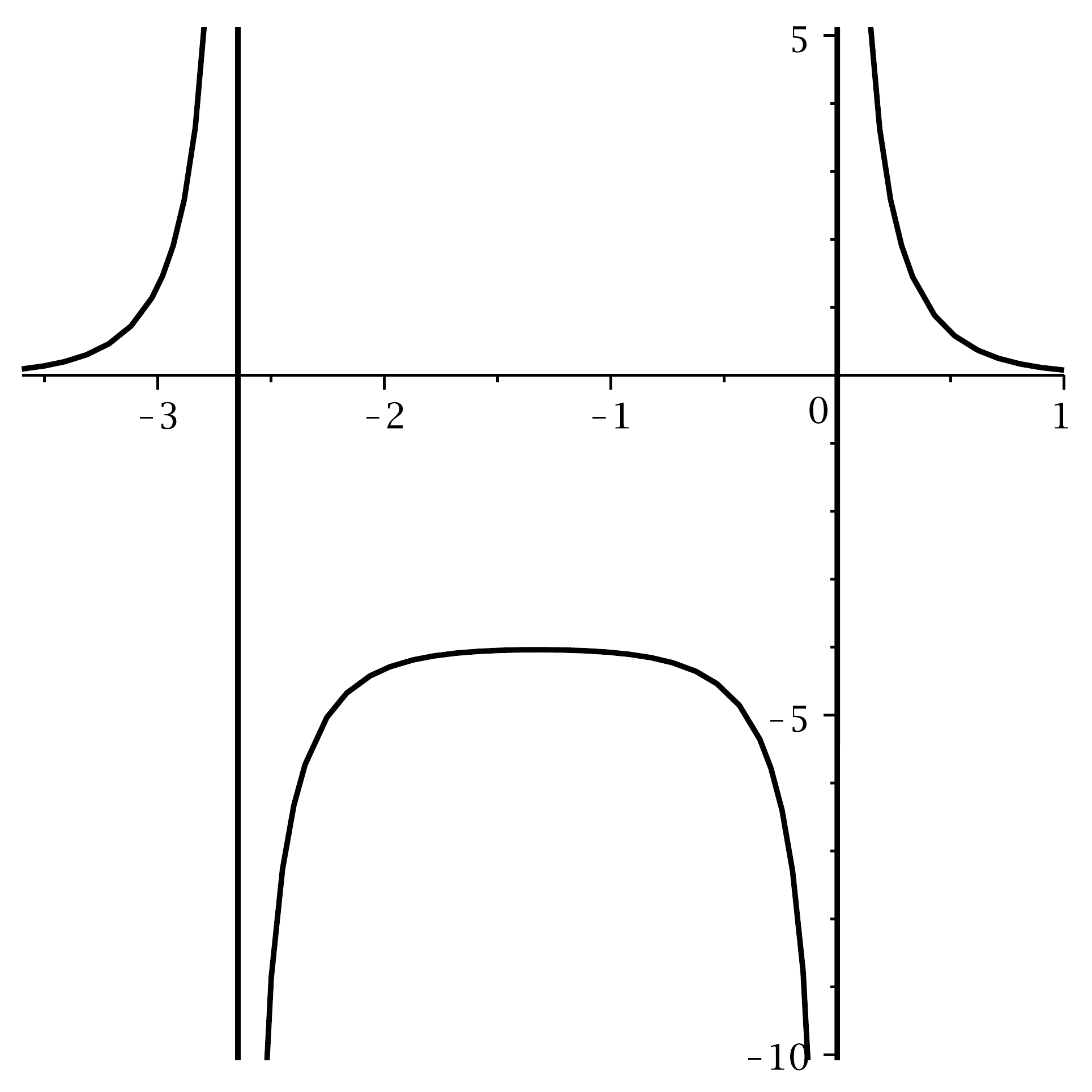}}
\subfloat[]{\label{fig:2c}
\includegraphics[width=0.3\textwidth]{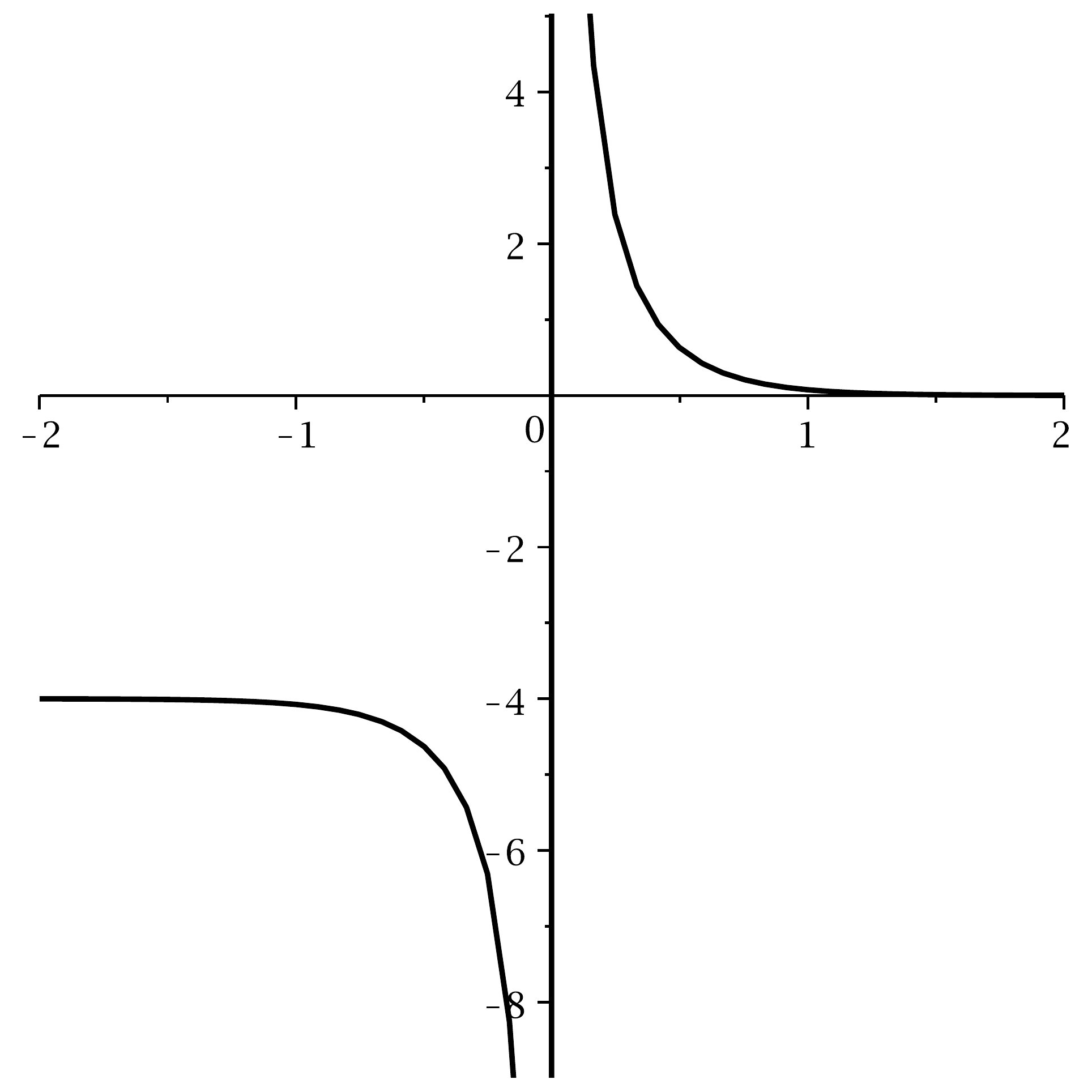}}
\caption{Soliton for the Gardner equation when $\k_1<0$ and 
$|\k_1| \le \mu$. $\mu=2$ and $\k_1 = (-1.5,\ -1.9999,\ -2)$ in 
figures (a, b, c), respectively.}
\label{fig_2}
\end{figure}

\begin{figure}
\centering
\subfloat[]{\label{fig_3a}
\includegraphics[width=0.3\textwidth]{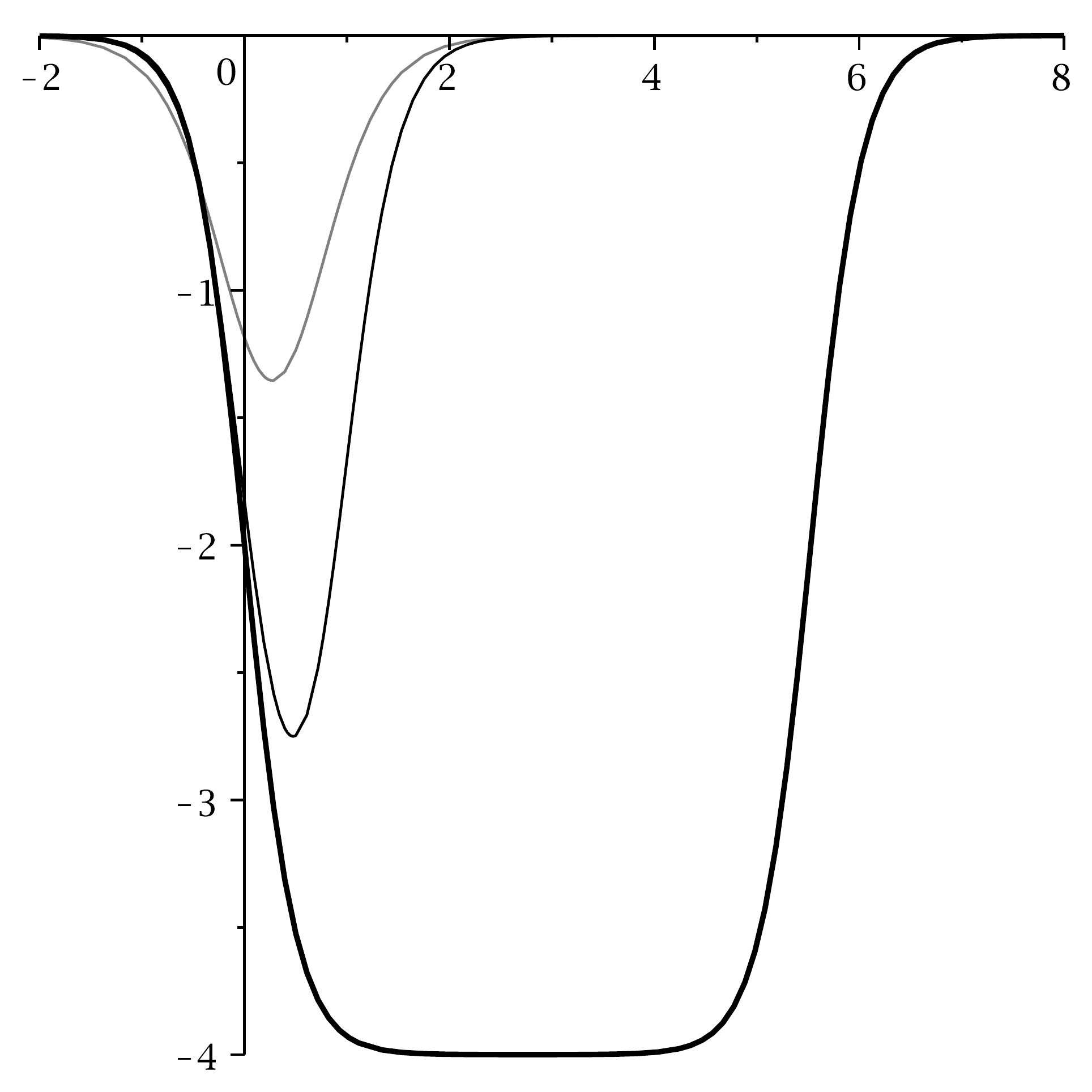}}
\subfloat[]{\label{fig_3b}
\includegraphics[width=0.3\textwidth]{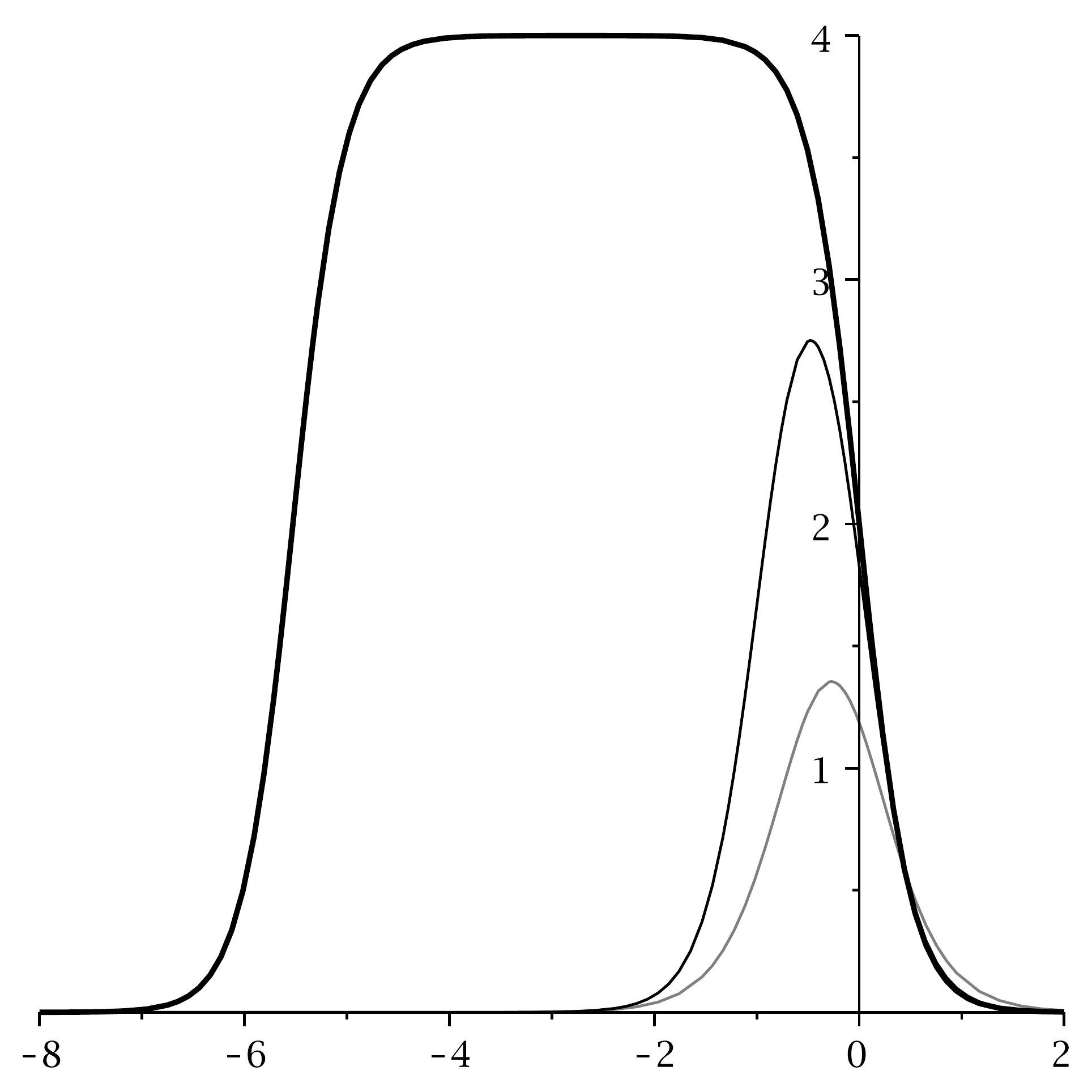}}
\subfloat[]{\label{fig_3c}
\includegraphics[width=0.3\textwidth]{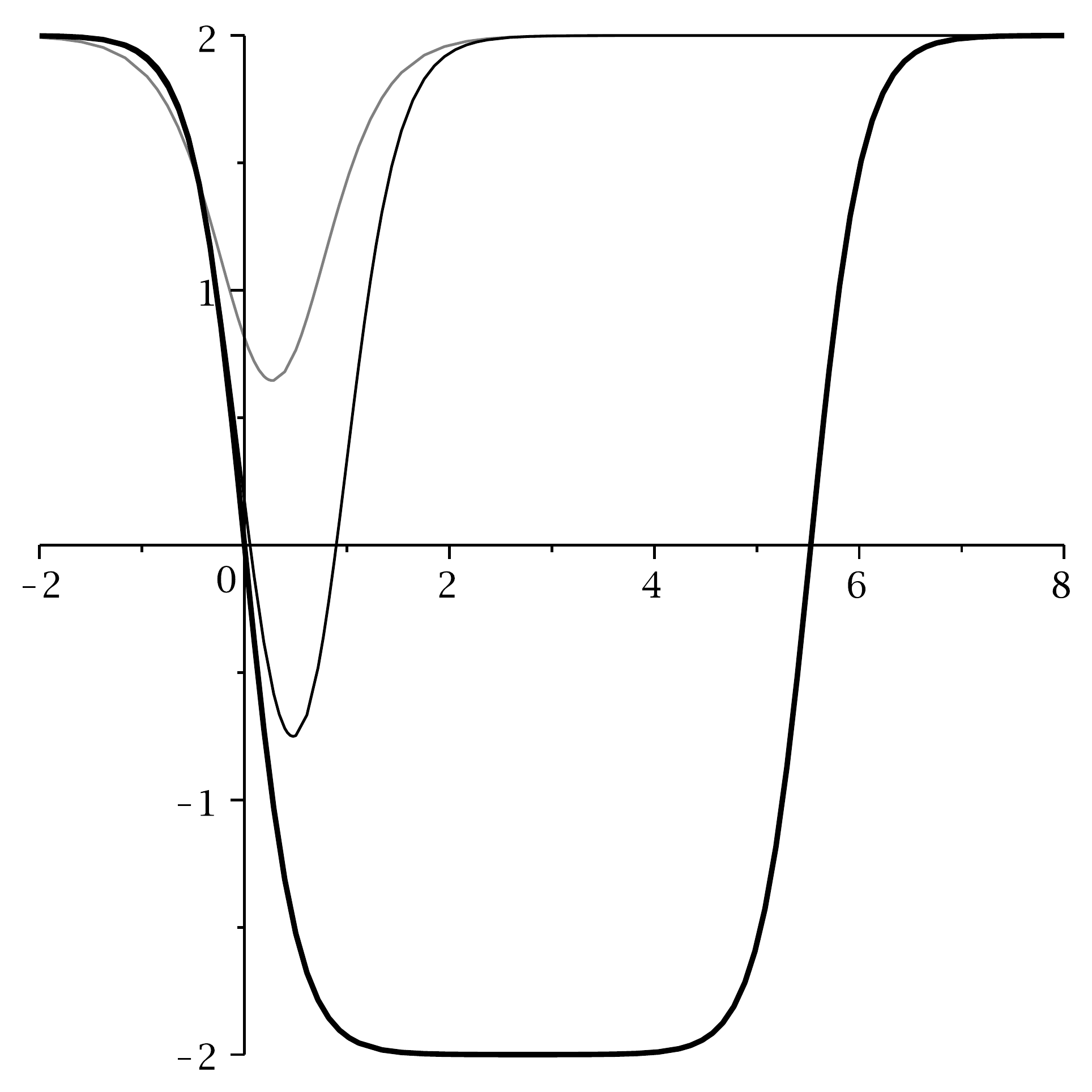}}
\caption{Soliton for Gardner (a, b) and mKdV (c) when 
$0<\k_1<\mu$.
The usual KdV like soliton increases its amplitude and becomes narrower
as $\k_1$ increases, but when $\k_1\to\mu^{-}$ it becomes the maximum
amplitude table-top soliton. $\mu=2$ and $\k_1=(1.5,\ 1.9,\ 1.999999999)$ in
(a). In (b) we changed sign $\mu\to-\mu, \; \k_1\to-\k_1$ with the same
numerical values. In (c) we have mKdV
dark solitons and table-top soltion, with the same numerical values as in (a).
Note the background corresponding to $v_0=2$.}
\label{fig_3}
\end{figure}

The \soliton{1} solution of the Gardner equation,
with dispersion \eref{spacetime}, has
different behaviour according to the values of $\k_1$ and $\mu$.
When $\k_1 \ge \mu $ and $\mu > 0$ the solutions are plotted in \fref{fig_1}.
Note that in \fref{fig_1c} we have a kink (anti-kink if you are
used to sine-Gordon terminology). 
When $\k_1 < 0$ but
$|\k_1|\le\mu$ we have \fref{fig_2}.  For $|\k_1| > \mu$ we recover
the behaviour of \fref{fig_1a}. The more interesting solution
occurs when $0 < \k_1 < \mu$, having the usual KdV solitons and also
a maximum amplitude table-top soliton, as shown in \fref{fig_3}.
The same graphs apply to \eref{gardner5}, 
the only difference is in the dispersion relation \eref{spacetime2}. 
If we change sign $\mu\to-\mu$, $\k_1\to-\k_1$ the
graphs are reflected through $y$ and $x$ axis, so we have an elevation  
instead of a depression wave, \fref{fig_3b}. In \fref{fig_3c} we have the
mKdV nonvanishing boundary solutions with dispersion \eref{spacetime-mkdv}.
The mKdV equation can also have a kink, like in \fref{fig_1a}, but
with the asymptotes in $+2$ and $-2$ corresponding to the constant
background $v_0=2$. The same solutions
also apply to \eref{msawada-kotera} with dispersion \eref{spacetime-mkdv5}.
This solutions with a constant background are known as dark solitons
\cite{chen,huang}. Dark solitons of the NLS equation has wide application 
in nonlinear optics.

Substituting the previous solutions of the mKdV hierarchy in the
Miura transformation \eref{miura} we obtain dark solitons of the KdV 
hierarchy. Unlike the mKdV, the KdV hierarchy does not have
kinks or table-top solitons, therefore, the interesting solutions
of the Gardner equation is inherited from the nonvanishing boundary 
solution of the mKdV equation.

\begin{figure}
\centering
\subfloat[]{\label{fig_4a}
\includegraphics[width=0.3\textwidth]{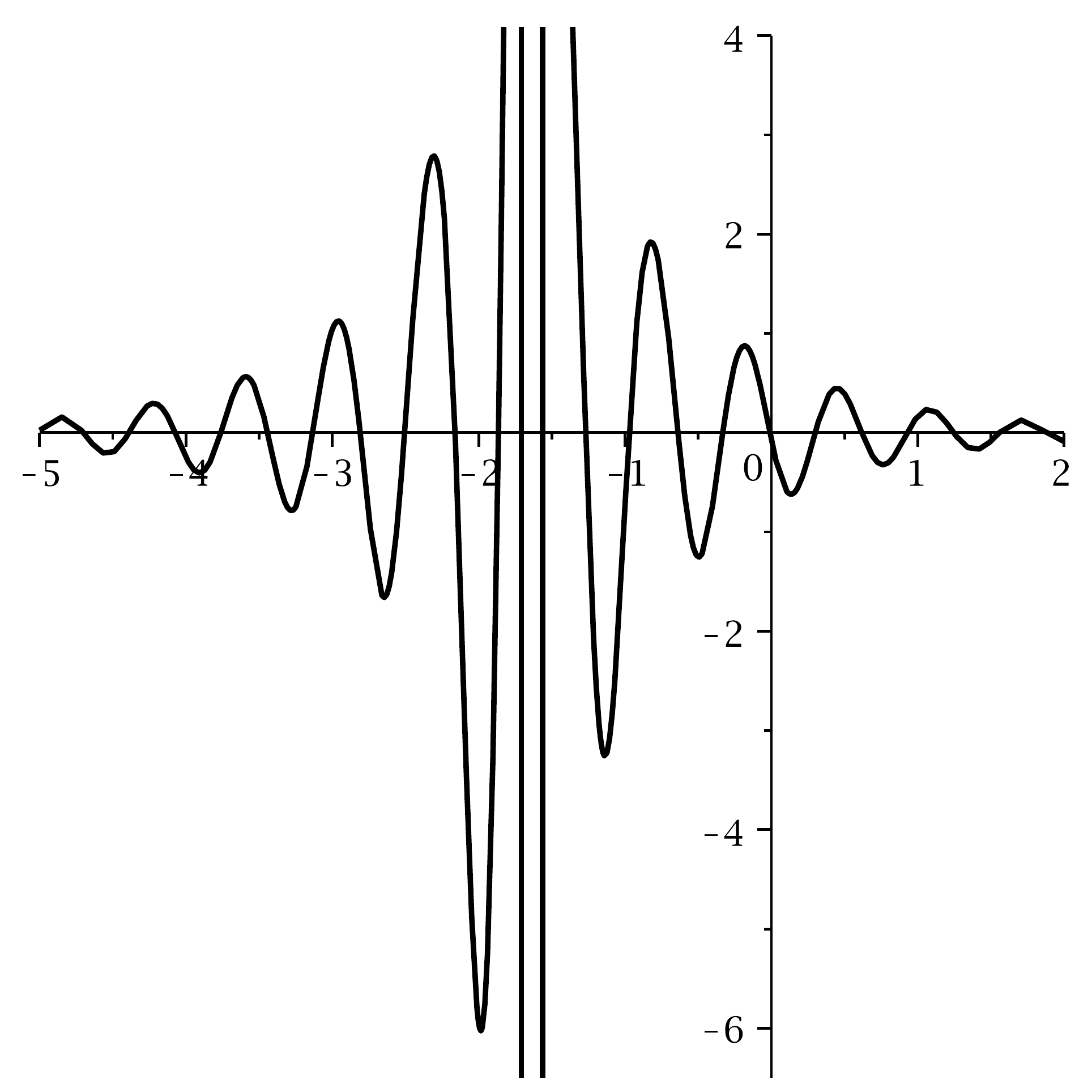}}
\subfloat[]{\label{fig_4b}
\includegraphics[width=0.3\textwidth]{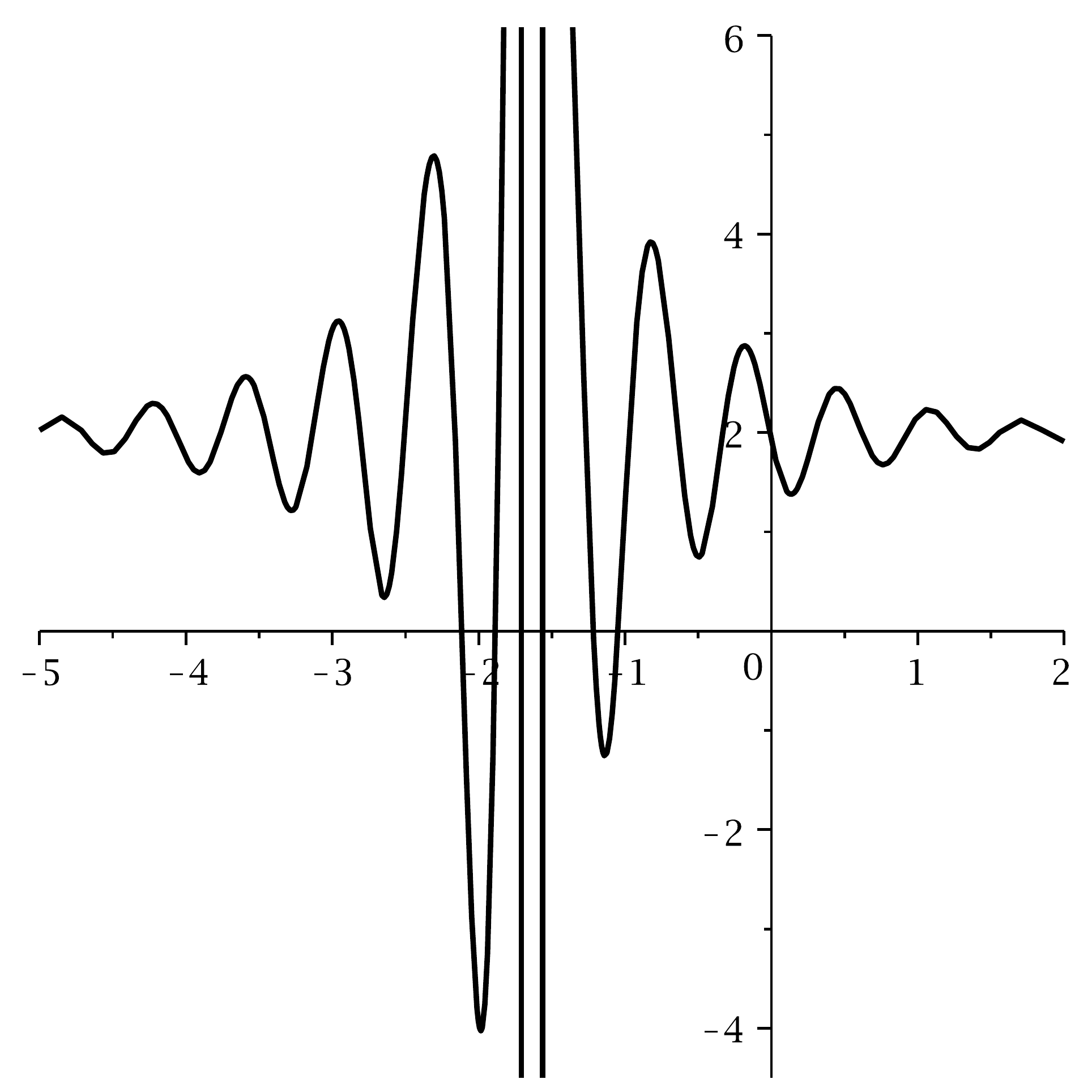}}
\subfloat[]{\label{fig_4c}
\includegraphics[width=0.3\textwidth]{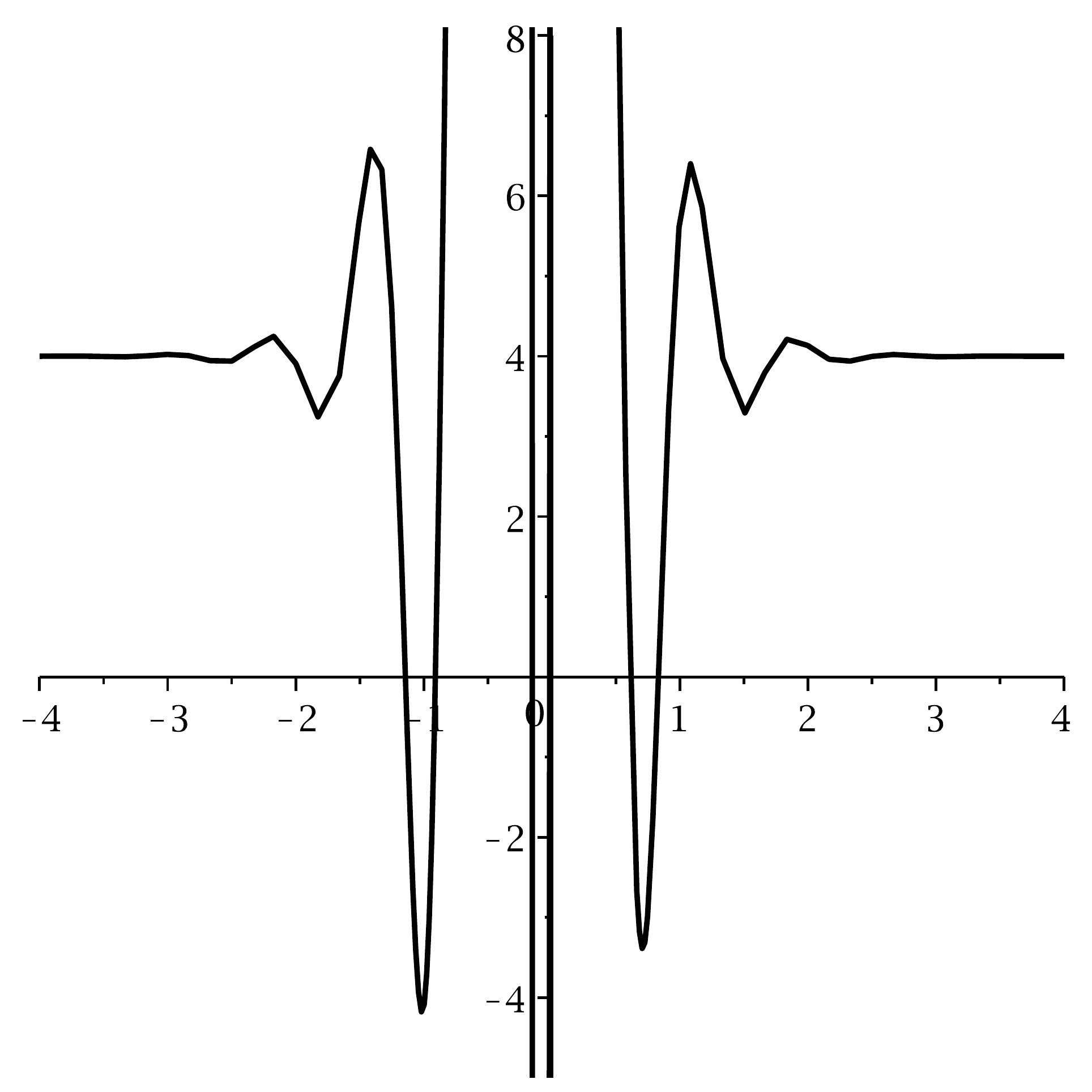}}
\caption{
(a) breather for the Gardner equation, (b) breather for the mKdV equation: 
$\mu=2$, $\a=0.5$, $\b=5$. 
(c) breather for the KdV equation: $\mu=2$, $\a=1.5$, $\b=4$.}
\label{fig_4}
\end{figure}

For the \soliton{2} solution \eref{2soliton} we can have a combination 
of two usual solitons, one soliton with a kink or one soliton with a 
table-top soliton. Moreover, we can have the interesting breather,
a spatially localized but oscillating solution, by choosing
complex conjugate wave numbers $\k_1=\a + i \b$ and $\k_2= \a -i \b$
(the explicit expression for the breather is given by \eref{tau_breather} 
in \ref{breather_wobble}).
\Fref{fig_4} shows the breather for the Gardner, mKdV and KdV equations.
The nonvanishing boundary breather of the KdV is obtained through 
Miura transformation. 

\begin{figure}
\centering
\subfloat[$t=5$]{\label{fig_5a}
\includegraphics[width=0.24\textwidth]{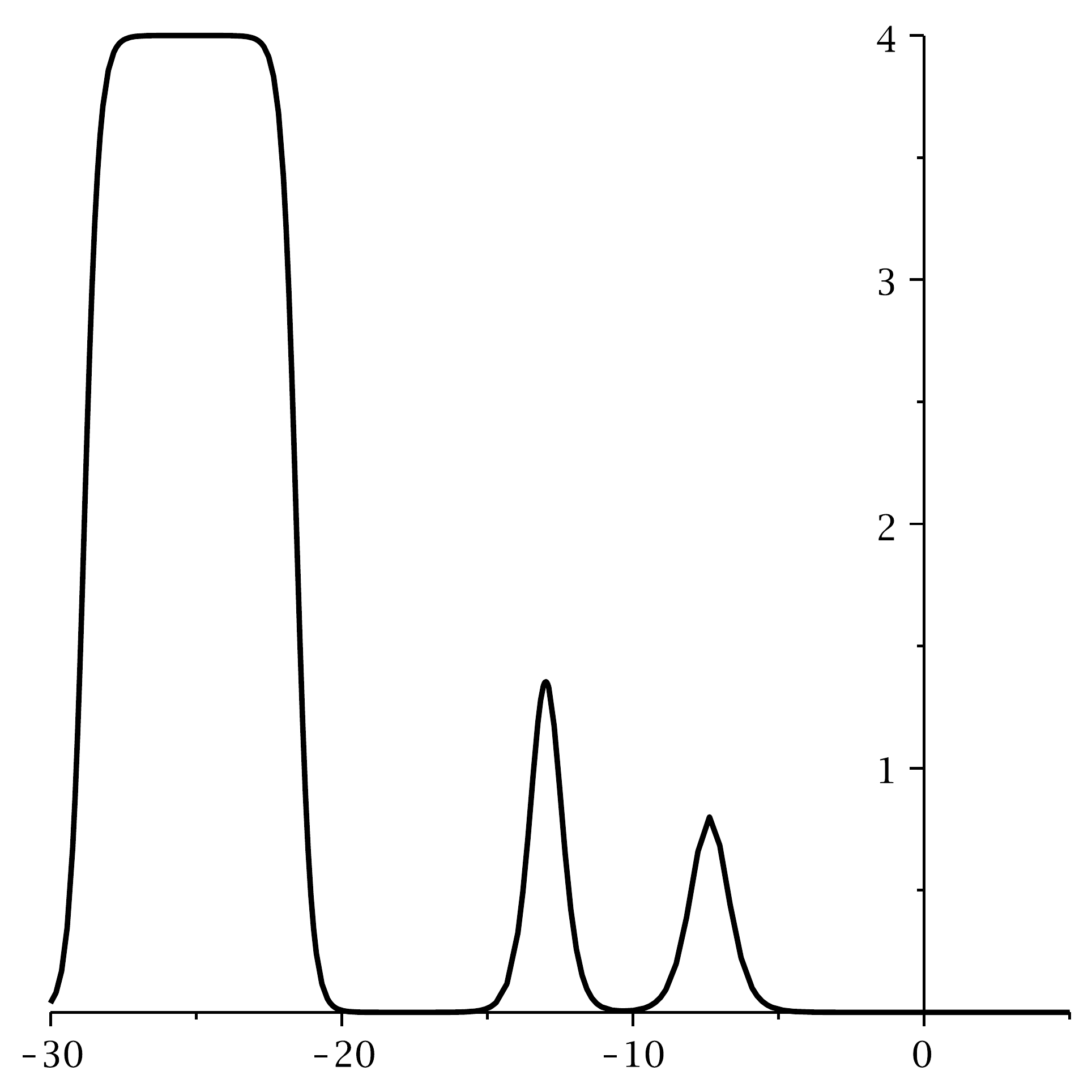}}
\subfloat[$t=2$]{\label{fig_5b}
\includegraphics[width=0.24\textwidth]{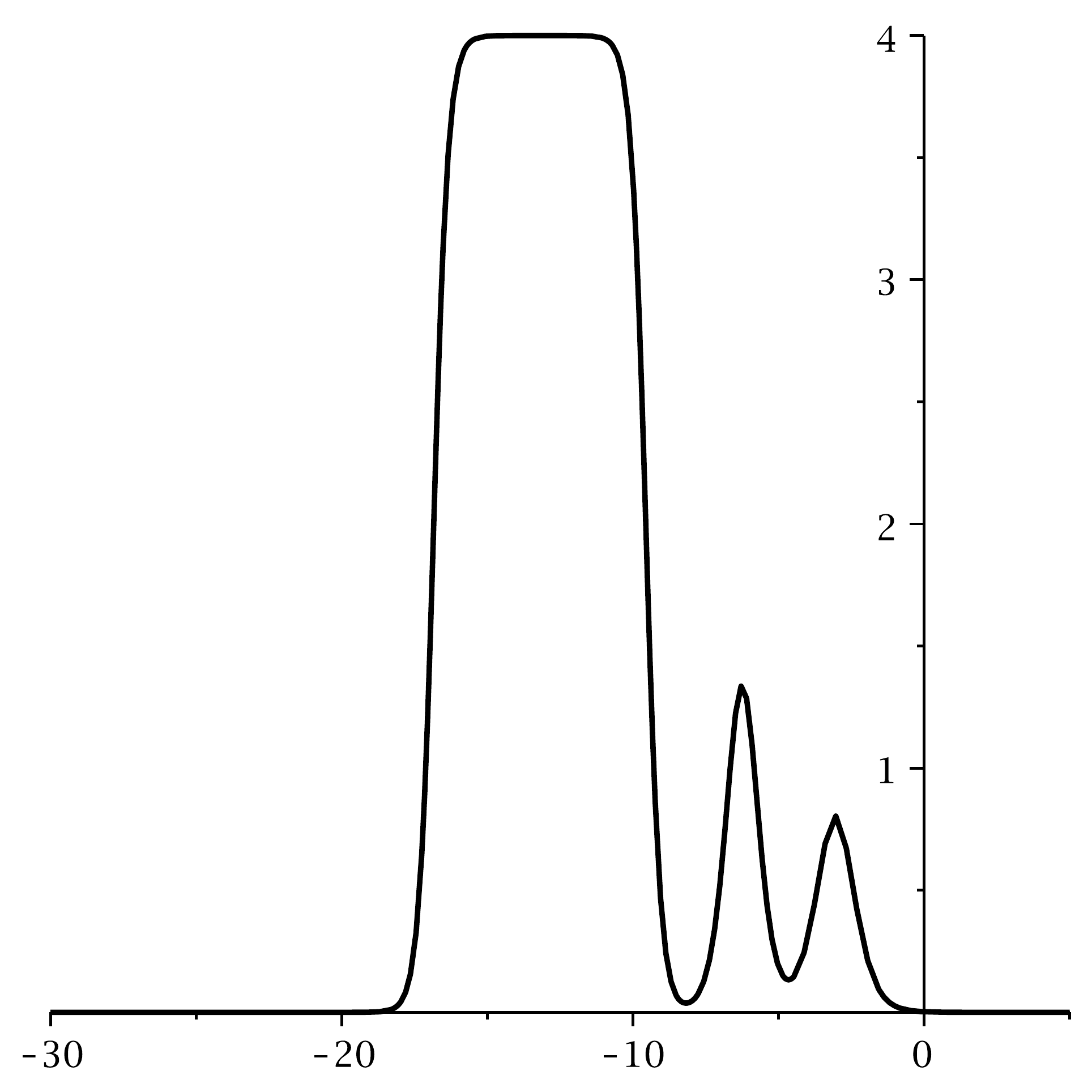}}
\subfloat[$t=0$]{\label{fig_5c}
\includegraphics[width=0.24\textwidth]{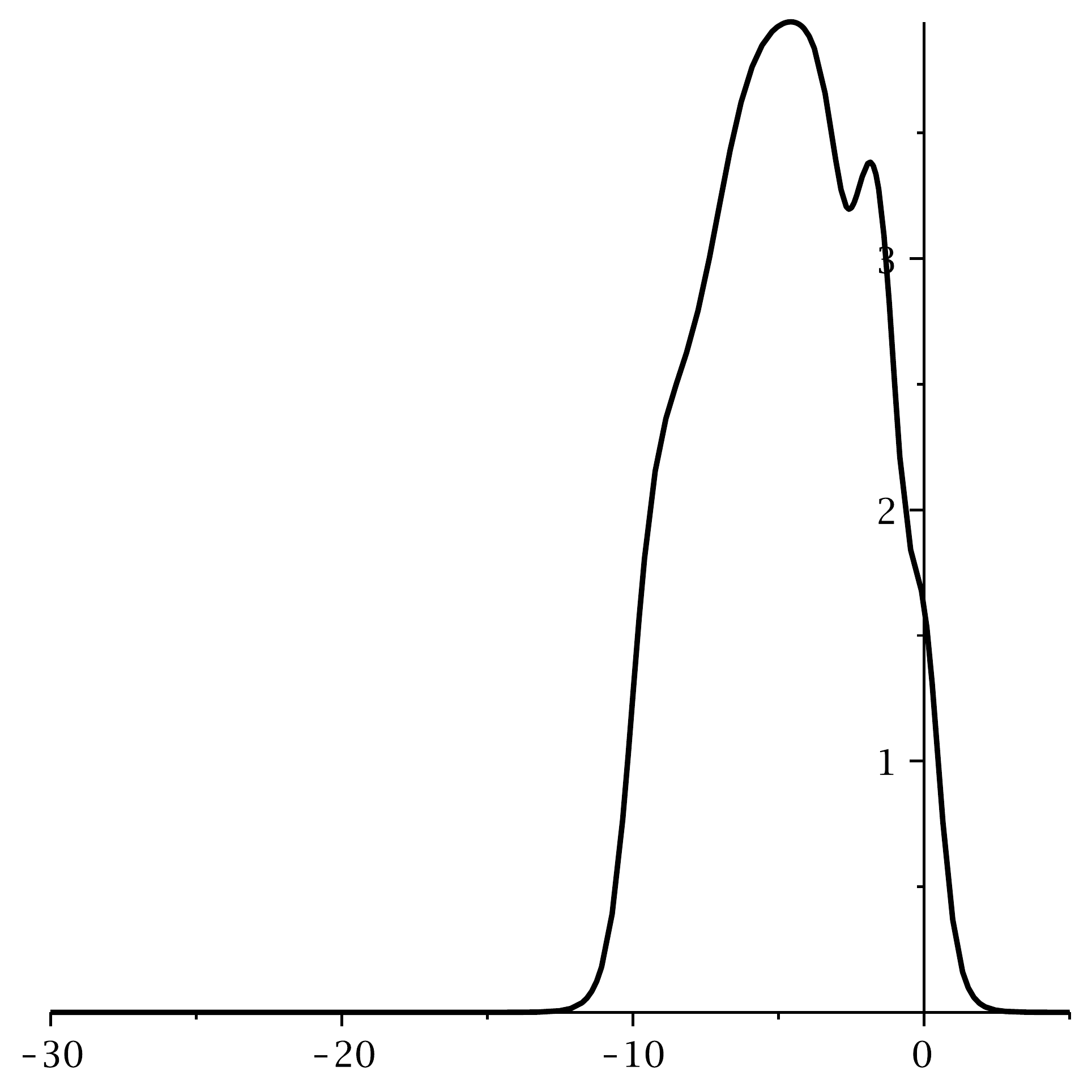}}
\subfloat[$t=-5$]{\label{fig_5d}
\includegraphics[width=0.24\textwidth]{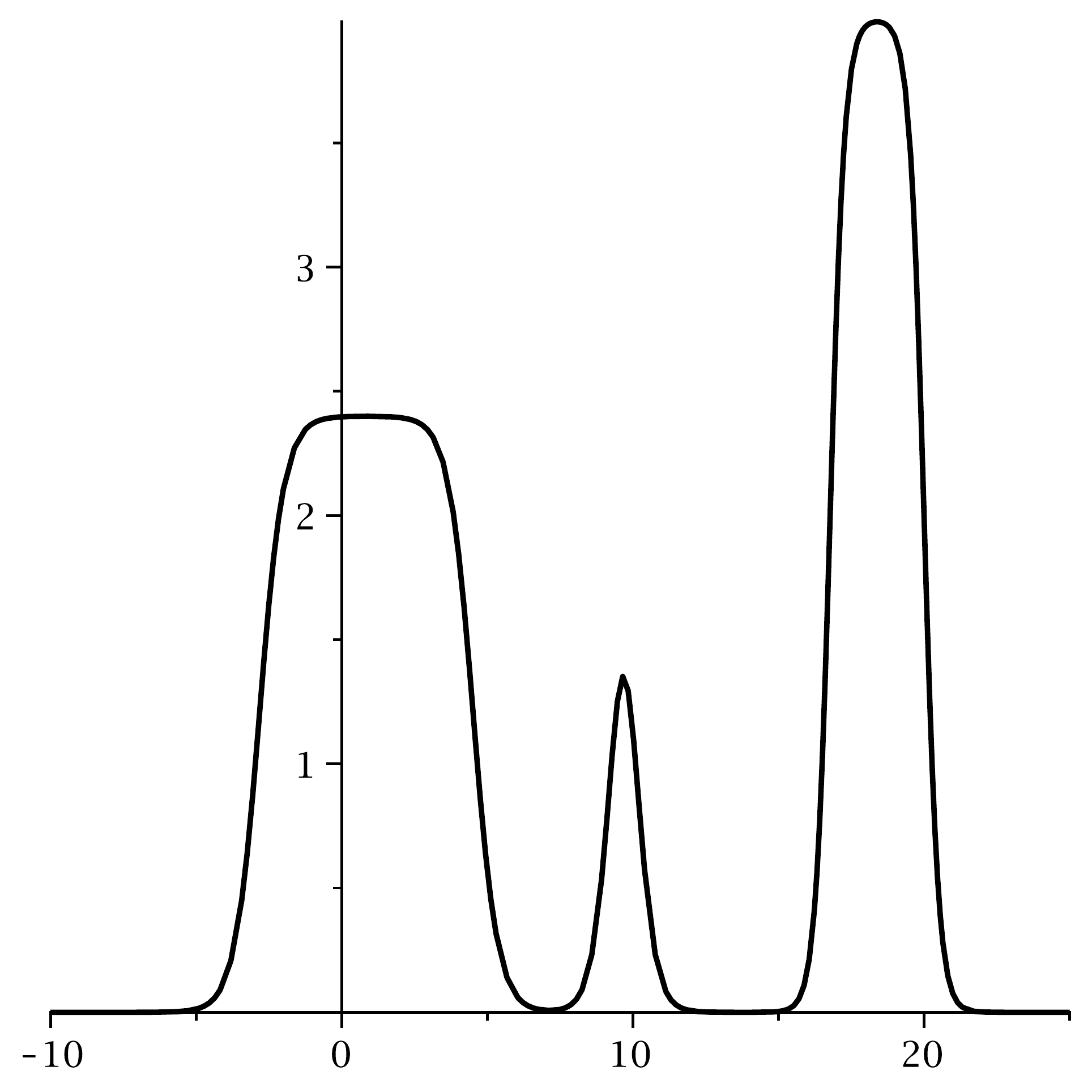}}
\caption{The waves travel to the left so you should read this 
figure from (d) to (a). We are plotting a 
\soliton{3} solution of the Gardner equation with parameters
$\mu=-2$, $\k_1=-1.99999$, $\k_2=-1.5$, $\k_3=-1.2$.}
\label{fig_5}
\end{figure}

In \fref{fig_5} we have a specific situation of the \soliton{3}
solution \eref{3soliton}. Note that the final profile is different
from the initial one. The waves did not keep their initial form
after interaction.
The mKdV equation can also have the same kind of \soliton{3} solution
with a background $v_0$.

\begin{figure}
\centering
\subfloat[]{\label{fig_6a}
\includegraphics[width=0.4\textwidth]{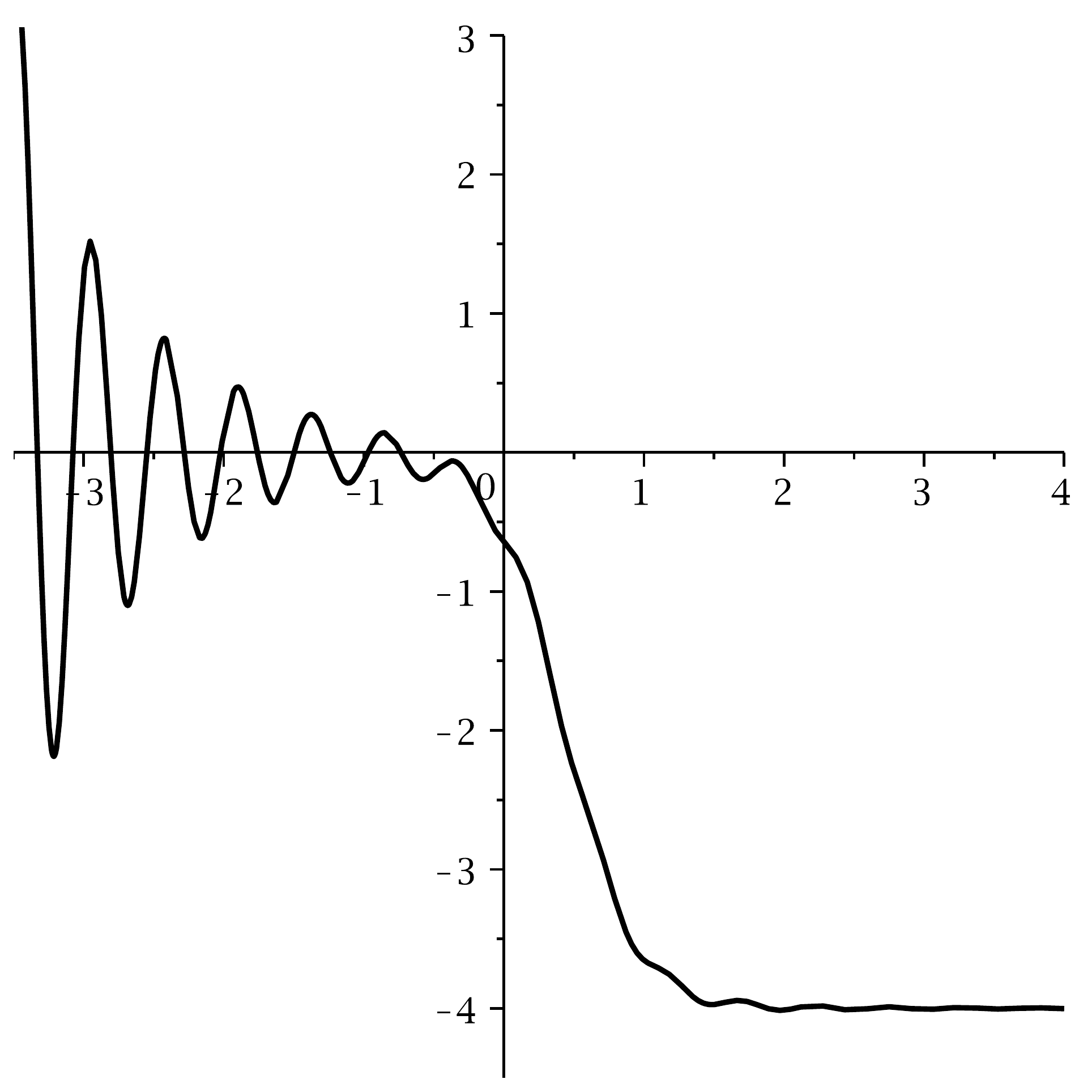}}
\subfloat[]{\label{fig_6b}
\includegraphics[width=0.4\textwidth]{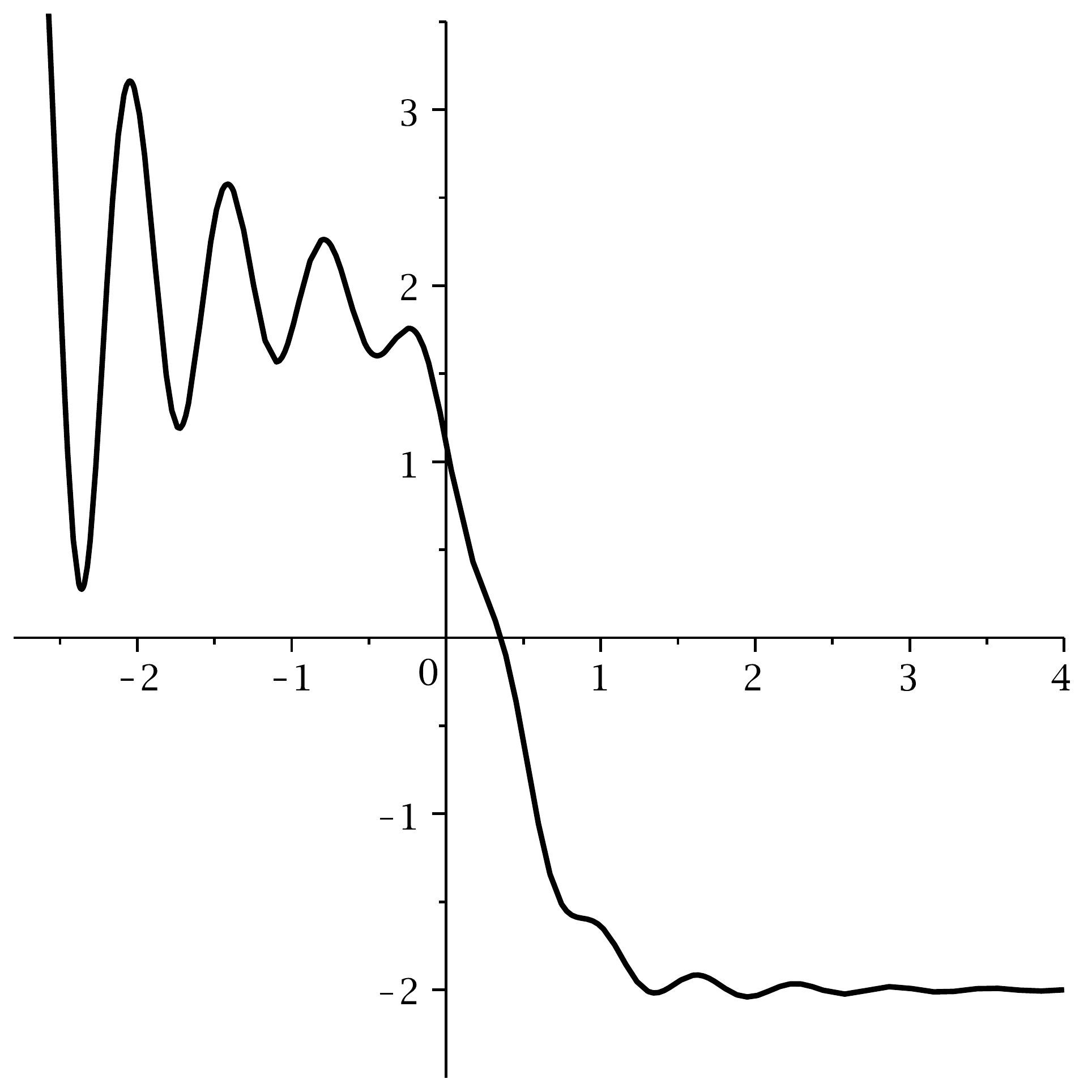}}
\caption{In (a) we have the wobble solution of the Gardner equation
with parameters
$\k_1=\frac{1}{2}+6i$, $\k_2=\frac{1}{2}-6i$, $\k_3=\mu=2$, $t=-0.02$. 
In (b) we have wobble for the mKdV equation:
$\k_1=\frac{1}{2}+5i$, $\k_2=\frac{1}{2}-5i$, $\k_3=v_0=2$, $t=-0.02$.}
\label{fig_6}
\end{figure}

In \cite{kalberman} K\" albermann proposed the wobble solution for the
sine-Gordon equation then Ferreira \etal \cite{luiz}
showed that the wobble is a \soliton{3} solution, where
two solitons combine to form a breather and the third one is a kink. 
We follow this line of thought and take our \soliton{3} solution 
\eref{3soliton} with 
$\k_1=\a+i\b$, $\k_2=\a-i\b$ and $\k_3=\mu$ to obtain the wobble solution 
for the Gardner equation (the explicit expression is \eref{tau_wobble} in 
\ref{breather_wobble}). 
We can also have the wobble for the mKdV equation
with $\k_3=v_0$. This solutions are shown in \fref{fig_6}. Despite
the KdV equation can have a breather, it can not have a wobble because it 
does not have a kink. We could also combine the breather with a table-top
soliton and with usual solitons as Grimshaw \etal 
considered in \cite{grimshaw2}.

\section{Concluding remarks}

We have constructed an integrable hierarchy \eref{gardner-hierarchy} that 
contains the Gardner equation \eref{gardner} as one of its members. 
This construction is based on the  
Kac-Moody algebra $\hat{s\ell}_2$ with principal gradation. 
Besides the Gardner equation, another fifth order PDE 
\eref{gardner5} that is a combination of Sawada-Kotera and fifth order 
mKdV \eref{msawada-kotera} was explicitly considered.

We have introduced a new vertex operator \eref{vertex} that used
within the dressing method, enabled us to obtain explicit \soliton{n}
solutions of the Gardner and mKdV hierarchies, this last one with 
nonvanishing boundary condition.
Besides usual KdV like solitons our solutions contemplate
table-top solitons and kinks. The \soliton{2} solution can be used to form a 
breather and using the Miura transformation we obtained a breather
for the KdV equation, also with nonvanishing boundary condition.

The \soliton{3} solution showed that the interaction of individual
waves when a kink or table-top soliton is present changes
its  initial profile such that  energy is always conserved. Combining
a breather with a kink we also obtained the wobble solution for the Gardner
and mKdV hierarchies.

We stress that the our solutions are valid for the whole hierarchy of nonlinear
equations, the only modification relying on the dispersion relations.

Further exploration of our results can be made by considering the
nonvanishing boundary value problem for the non-abelian AKNS hierarchy that
contains the NLS equation which has practical applications in nonlinear
optics and water waves.

\ack
We thank CNPq and Fapesp for support. We also thank the anonymous referees
for valuable suggestions.

\appendix

\section{Breather and wobble solutions}\label{breather_wobble}

The general breather expression is obtained from \eref{2soliton} 
by setting complex conjugate wave numbers, $\k_1=\k_2^{*}=\a+i\b$,
which implies that the dispersion relations will be in the form
$\xi_1=\xi_2^{*}=\eta + i\zeta$, for some real functions $\eta=\eta(x,t)$
and $\zeta=\zeta(x,t)$ to be determined latter.
From \eref{coeff1} we have $\expected{V_1}_j=\expected{V_2}_j^{*}=a_j-ib$ 
where 
\br
a_j &= \frac{\mu\a}{2\(\a^2+\b^2\)}+\frac{\sigma_j}{2},
\qquad (\sigma_0=-1,\;\sigma_1=1), \label{coeff_aj} \\
b &= \frac{\mu\b}{2\(\a^2+\b^2\)},\label{coeff_b}
\er
and from \eref{coeff2}
\be
\expected{V_1V_2}_j=-\frac{\b^2}{\a^2}\(a_j^2+b^2\).
\label{coeff_v1v2}
\ee
Therefore, the general tau functions for the breather are
\be
\tau_j = 1+2e^\eta\lb a_j\cos \zeta  + b\sin \zeta - 
\frac{\b^2}{2\a^2}\(a_j^2+b^2\)e^{\eta}\rb. \label{tau_breather}
\ee

The wobble is obtained from \eref{3soliton} with $\k_1=\k_2^{*}=\a+i\b$
and $\k_3=\mu$, so $\xi_1=\xi_2^{*}=\eta + i\zeta$ and
$\xi_3\equiv\eta_\mu$ is a real function depending on $\mu$.
From \eref{coeff1} we get $\expected{V_3}_j=\delta_{j1}$ and 
from \eref{coeff2} we calculate  
$\expected{V_1V_3}_j=\expected{V_2V_3}_j^{*}=\delta_{j1}\(c_j+id_j\)$
where
\br
\frac{\k_1-\k_3}{\k_1+\k_3} = 
\(\frac{\k_2-\k_3}{\k_2+\k_3}\)^{*} = \gamma + i\nu, \\
\gamma = \frac{\a^2+\b^2-\mu^2}{\(\a+\mu\)^2+\b^2}, \qquad
\nu = \frac{2\b\mu}{\(\a+\mu\)^2+\b^2},
\label{gamma_nu}
\er
and
\br
c_j = a_j\(\gamma^2-\nu^2\)+2b\gamma\nu, \label{coeff_cj} \\
d_j = 2a_j\gamma\nu-b\(\gamma^2-\nu^2\). \label{coeff_dj}
\er
The general wobble tau functions are then given by
\br
\fl \t_j &= 1+2e^{\eta}\lb a_j\cos\zeta + b\sin\zeta - 
\frac{\b^2}{2\a^2}\(a_j^2+b^2\)e^\eta \rb + \nonu \\
\fl &+2\delta_{j1}e^{\eta+\eta_\mu}\lb c_j\cos\zeta - d_j\sin\zeta
-\frac{\b^2}{2\a^2}\(a_j^2+b^2\)\(c_j^2+d_j^2\)e^\eta + 
\frac{1}{2}e^{-\eta} \rb.
\label{tau_wobble}
\er

\subsection{Gardner hierarchy solutions}

The breather or wobble of the Gardner equation \eref{gardner} is given by 
\be
v = \pa_x\ln\frac{\t_0}{\t_1}
\label{v_gardner}
\ee
replacing \eref{tau_breather} or \eref{tau_wobble}, respectively. 
From \eref{spacetime} we have
\br
\eta &= 2\a x+2\a\(\a^2-3\b^2\)t, \\
\zeta &= 2\b x-2\b\(\b^2-3\a^2\)t, \\
\eta_\mu &= 2\mu x + 2\mu^3 t.
\er
For \eref{gardner5} the only modification comes from \eref{spacetime2} 
that yields
\br
\eta &= 2\a x+2\a\(\a^4-10\a^2\b^2+5\b^4\)t, \\
\zeta &= 2\b x+2\b\(\b^4-10\a^2\b^2+5\a^4\)t, \\
\eta_\mu &= 2\mu x + 2\mu^5 t.
\er

\subsection{mKdV hierarchy solutions}

For the mKdV equation \eref{mkdv}, the breather or wobble is given by
\be
v = v_0 + \pa_x\ln\frac{\t_0}{\t_1}
\label{v_mkdv}
\ee
using \eref{tau_breather} or \eref{tau_wobble}, respectively. 
In the coefficients \eref{coeff_aj}, \eref{coeff_b}, \eref{coeff_cj}
and \eref{coeff_dj} we should replace $\mu \to v_0$. 

Taking into account the dispersion \eref{spacetime-mkdv} we have
\br
\eta &=2\a x + 2\a\(\a^2-3\b^2-\case{3}{2}v_0^2\)t, \\
\zeta &=2\b x - 2\b\(\b^2-3\a^2+\case{3}{2}v_0^2\)t, \\
\eta_{v_0} &= 2v_0 x - v_0^3 t.
\er
For \eref{msawada-kotera} the dispersion \eref{spacetime-mkdv5} implies
\br
\eta &=2\a x + 2\a\(\a^4-10\a^2\b^2+5\b^4-\case{5}{2}v_0^2\a^2
+\case{15}{2}v_0^2\b^2+\case{15}{8}v_0^4\)t, \\
\zeta &=2\b x + 2\b\(\b^4-10\a^2\b^2+5\a^4+\case{5}{2}v_0^2\b^2
-\case{15}{2}v_0^2\a^2+\case{15}{8}v_0^4\)t, \\
\eta_{v_0} &= 2v_0 x + \case{3}{4}v_0^5 t.
\er


\end{document}